\documentclass[journal=jctce,manuscript=article]{achemso}
\usepackage[version=3]{mhchem}

\usepackage{graphicx} 
\usepackage{subfig}
\usepackage{multirow}
\usepackage{longtable}
\usepackage{achemso} 
\usepackage{array,booktabs}
\usepackage{float}
\newcolumntype{C}[1]{>{\centering\arraybackslash}p{#1}}
\usepackage{caption}
\usepackage{color}
\usepackage[]{siunitx}
\usepackage{verbatim} 
\usepackage{algorithm2e}
\setkeys{acs}{usetitle = true}
\usepackage{color, colortbl}
\definecolor{Gray}{gray}{0.9}

\author{Neha$^\dag$}
\affiliation{Department of Chemistry, Indian Institute of Technology, Delhi,\\ Hauz Khas, New Delhi 110016, India}
\author{Vikas  Tiwari$^\dag$}
\affiliation{Department of Chemistry, Indian Institute of Technology, Delhi,\\ Hauz Khas, New Delhi 110016, India}
\author{Soumya Mondal$^\dag$}
\affiliation{Department of Chemistry, Indian Institute of Technology, Delhi,\\ Hauz Khas, New Delhi 110016, India}
\author{Nisha Kumari$^\dag$}
\affiliation{Department of Chemistry, Indian Institute of Technology, Delhi,\\ Hauz Khas, New Delhi 110016, India}
\author{Tarak Karmakar}
\affiliation{Department of Chemistry, Indian Institute of Technology, Delhi,\\ Hauz Khas, New Delhi 110016, India}
\email{tkarmakar@chemistry.iitd.ac.in}

\title[\texttt{achemso}]
{Collective Variables for Crystallization Simulations - from Early Developments to Recent Advances}

\begin{document}
\footnote{$\dag$ Authors have contributed to the review equally.}
\onecolumn
\maketitle
\begin{abstract}
{Crystallization is one of the most important physicochemical processes which has relevance in material science, biology, and the environment. Decades of experimental and theoretical efforts have been made to understand this fundamental symmetry-breaking transition. While experiments provide equilibrium structures and shapes of crystals, they are limited to unraveling how molecules aggregate to form crystal nuclei that subsequently transform into bulk crystals. Computer simulations, mainly molecular dynamics (MD), can provide such microscopic details during the early stage of a crystallization event. Crystallization is a {\em rare event} that takes place in timescales much longer than a typical equilibrium MD simulation can sample. This inadequate sampling of the MD method can be easily circumvented by the use of enhanced sampling (ES) simulations. An ES method enhances the fluctuations of a  system's slow degrees of freedom, called collective variables (CVs), by applying a bias potential, and thereby transforms the system from one state to the other within a short timescale. The most crucial part of an ES method is to find suitable CVs, which often needs intuition and several trial-and-error optimization steps. Over the years, a plethora of CVs has been developed and applied in the study of crystallization. In this review, we provide a brief overview of CVs that have been developed and used in ES simulations to study crystallization from melt or solution. These CVs can be categorized mainly into four types: (i) spherical particle-based, (ii) molecular template-based, (iii) physical property-based, and (iv) CVs obtained from dimensionality reduction techniques. We present the context-based evolution of CVs, discuss the current challenges, and propose future directions to further develop effective CVs for the study of crystallization of complex systems.}
\end{abstract}

\newpage 
\noindent
\section{1. INTRODUCTION}

Understanding phase transition using computer simulations has been a prime focus of the simulators - starting from the early method developers to the the current community. Alongside the enrichment of our fundamental understanding of the symmetry-breaking transition from a non-symmetric liquid state to a symmetric crystalline phase, this process is of great interest and bears relevance to the pharmaceutical industry as well as the environment. Investigation of the crystallization mechanism, optimizing crystallization conditions, predicting crystal shape, and obtaining thermodynamics and kinetics of crystal growth/dissolution processes require significant time investment as well as monetary resources.

Computer simulations, especially the molecular dynamics (MD) methods, have been probably the most convenient way to delineate microscopic details of the early stage of the crystallization process and calculate the thermodynamics and kinetics of the process. Unfortunately, in the context of computer simulations, crystallization is a rare event that, in most cases, takes place in a timescale ranging from milliseconds to seconds. The brute-force MD simulations are limited by short timescales in the range of nano- to microseconds that are inadequate to study crystallization. To circumvent this issue, several enhanced sampling (ES) simulations methods have been developed over the years.~\cite{Torrie1997,laio2002escaping,barducci2008well,valsson2014variational,valsson2016enhancing,invernizzi2020rethinking,debnath2020gaussian,invernizzi2020unified} The central aspect of most of these methods is to define one or more variables that are functions of atomic coordinates and describe the system's slow degrees of freedom. These variables are called order parameters (OPs) or collective variables (CVs) (note that there are subtle differences between these two nomenclatures). In ES simulations, the fluctuations of these CVs are enhanced to sample the metastable states. In the context of crystallization, the CVs are designed such a way that they can distinguish between the particles in the crystal and liquid or dispersed in solution phases.   

The CVs that are routinely used in crystallization simulations can be majorly divided into four categories, (i) spherical particle-based CVs such as the most popular Steinhardt parameters that are useful in simulating spherical particles (atomic, metallic, and colloids systems), (ii) molecular CVs (template-based, root mean square deviation-based, local crystallinity) that are used to crystallize molecular systems in pre-defined structures, (iii) physical property-based CVs such as volume/density, pair correlation functions, structure factor and X-ray diffraction peaks, and entropy-enthalpy, and (iv) CVs that are derived from linear and non-linear (machine learning) dimensionality reduction of basic OPs (distance, coordination number, angles, etc.). 

In this review, we provide a systematic overview of the CVs that are used in ES simulations of crystallization. In doing so, we might miss many other important developments including those CVs that are used as classifiers (fingerprints) to characterise disordered and various crystal polymorphs. Discussion on these topics can be found in the literature.~\cite{chau1998new,errington2003quantification,radhakrishnan2002new,hawtin2008gas,peters2009competing,angioletti2010solid,geiger2013neural,long2014nonlinear,cheng2015solid,lee2019entropic,fulford2019deepice,defever2019generalized} We conclude this review with a discussion on the current challenges and future directions in the development of efficient CVs for the study of crystallization of complex molecular systems.\\

\noindent
\section{2. COLLECTIVE VARIABLES (CVs)}

\subsection{2.1 Spherical-particle-based}
\noindent
\subsubsection{2.1.1. Steinhardt parameters}
In 1983, Steinhardt {\em et al.} proposed bond-orientational order parameters (OPs) to characterize and distinguish between solid and liquid states.~\cite{steinhardt1983bond} These OPs identify the system's states by measuring the symmetries of the clusters formed during simulation. In their approach, a central atom, $i$ is considered which forms {\em `virtual bonds'} with its neighbors that are found within the radial distance of $1.2r_0$ around it$i$, where $r_0$ is the minimum distance in the  Lennard-Jones (LJ) potential. Here these `{\em virtual bonds}' do not imply any chemical bonds rather they are imaginary lines that connect the central atom with its neighbors. The atoms' connectivity is defined by spherical harmonic function, and the OP, $\bar{q}_{lm}(\Vec{r})$ for particle $i$ is defined by taking an average of these spherical harmonics over a suitable set of neighbors around the particle $i$,

\begin{equation}
\bar{q}_{lm}(i) = \frac{1}{N_b(i)}\sum_{j=1}^{N_b(i)}  Y_{lm}( \theta(r_{ij}),\phi(r_{ij})) 
\label{eq:stein}
\end{equation}

Here $N_b(i)$ is number of virtual bonds of particles $i$ with its neighbors, $Y_{lm} (\theta,\phi)$ are the spherical harmonics, and $\theta(r_{ij})$ and $\phi(r_{ij})$ are polar angles made by bonds with respect to some reference coordinate system. The value of  $Y_{lm} (\theta,\phi)$ depends upon two intergers $l$ and $m$ and for a given value of l there are $2l + 1$ values of $m$. Hence, $q_{lm}(i)$ will have different values  for a particular value of $l$. Therefore a particular structure will have different values of $q_{lm}(i)$ for a particular value of $m$ for a given $l$. Therefore rotationally invariant combinations of bond order parameters are defined as:

\begin{align*}
   q_l(i) &=\sqrt{\frac{4\pi}{2l+ 1} \sum_{m=-l}^{l}|{\bar{q}_{lm}(i)}|^2} \\
   \hat{W_l} &= \frac{W_l}{(\sum_{l=-m}^{m}|\bar{q}_{lm}|^2)^\frac{3}{2}}
\end{align*}
where, 
\begin{equation}
W_l =\sum_{\substack{m_1,m_2,m_3 \\ m_1 +m_2 +m_3 =0 }}\begin{pmatrix}
l & l & l\\
m_1 & m_2 & m_3
\end{pmatrix} \bar{q}_{lm_1} \bar{q}_{lm_2}\bar{q}_{lm_3}
\end{equation}

\begin{figure}[H]
    \centering
    \includegraphics[width=0.6 \linewidth]{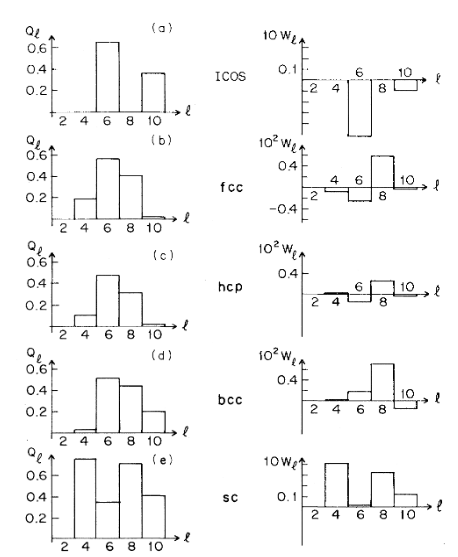}
    \caption{Histogram of $q_{lm}$ and $W_l$ for 13 atom icosahedral, fcc, and hcp clusters, as well for 15 atom bcc and 7 atom sc cluster. {(\em Reproduced
from Ref. \citenum{steinhardt1983bond}, with the permission of APS Publisher.)}}
    \label{fig:stein}
\end{figure}

The matrix 
$\begin{pmatrix}
l & l & l\\
m_1 & m_2 & m_3
\end{pmatrix} $ is called Wigner $3j$ symbols. The $q_{lm}$ and $W_{l}$ are called as quadratic and third order invariants, respectively. Figure \ref{fig:stein} shows the histogram of $q_{lm}$ and $W_{l}$ for five clusters at $l=2, 4, 6, 8, 10$. Non-zero averages appear at $l\geq 4 $ for clusters having hcp and cubic symmetry. For icosahedral clusters, non-zero averages occur at $l=6$.  All the values are calculated for clusters corresponding to the unit cell.

In 1996, ten Wolde {\em et al.}~\cite{rein1996numerical} studied the crystallization of LJ system using MD simulations. They have computed nucleation barrier, nucleation rate, and identified pre-critical, critical and post-critical nuclei for LJ system. They have calculated the free energy surface for nucleation by using the umbrella sampling method.\cite{Torrie1997} In order to calculate the nucleation barrier, they defined an OP (reaction coordinate) which measures the degree of crystallinity of a system during phase transition. They found that the local OPs introduced by Steinhardt (Eq. \ref{eq:stein}) has almost the same values in both solid and liquid phases. Therefore, they introduced the global Steinhardt parameters, $q_6$ which vanishes in liquid and have high values in the crystalline phase. The generalized global orientational OP, $q_{lm}$ is defined as:

\begin{equation}
\bar{q}_{lm}=\frac{\sum_{i=1}^{N}N_b(i)\bar{q}_{lm}(i)}{\sum_{i=1}^{N} N_b(i)}
\end{equation}
In $\bar{q}_{lm}$, the average is taken over all $N$ particles present in the system. The values of global orientational OPs for different crystal systems  can be seen in the Table \ref{fig:TAB1},

\begin{center}
\captionof{table}{Global orientational order parameters for fcc, hcp, bcc, sc, and icosahedral structures. {(\em Reprinted with the permission of AIP publisher from ref. \citenum{rein1996numerical})}}
\begin{tabular}{  m{4cm} m{2cm} m{2cm}   m{2cm}  m{2cm}  }
\hline \hline
  & $Q_4$ & $Q_6$ & $ \Hat{W}_4$ & $ \Hat{W}_6$  \\
\hline
fcc & 0.191 & 0.575 & -0.159 & -0.013  \\ 
hcp & 0.097 & 0.485 &0.134 & -0.012 \\ 
bcc & 0.036 & 0.511 & 0.159 & 0.013 \\ 
sc & 0.764 & 0.354 & 0.159 & 0.013\\
Icosahedral & 0 & 0.663 & 0 & -0.170\\
liquid & 0 & 0 & 0 & 0\\
\hline
\end{tabular}
\label{fig:TAB1}
\end{center}

Subsequently, ten Wolde, Frenkel and coworkers~\cite{ten1995numerical} introduced the scalar product of the normalized bond orientational vectors, $\bf{q_l(i)}$ and $\bf{q_l(j)}$ between neighboring particles $i$ and $j$,  used to differentiate between solid and liquid clusters as described below :
\begin{equation}
 \bf{q_l(i)}.\bf{q_l(j)} =\sum_{m=-l}^{l}\bf{q_l(i)}.\bf{q_l(j)}^*
\label{eq:dotq}
\end{equation}

\begin{figure}[H]
    \centering
    \includegraphics[width=0.6 \linewidth]{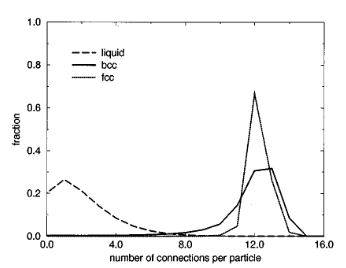}
    \caption{Distribution of number of connections per particles for liquid, bcc, and fcc structure. {(\em Reprinted with the permission of AIP publisher from Ref. \citenum{rein1996numerical})}}
    \label{fig:wolde}
\end{figure}

Here, $\bf{q_l(j)^*}$ is the complex conjugate of $\bf{q_l(j)}$, and $\bf{q_l(i)}.\bf{q_l(i)}$ is equal to 1. Two neighbors, $i$ and $j$ are connected if the {\em dot product} is greater than a threshold, say 0.5. According to this criterion all particles in solid state are found to be connected to each other. However this is not a sufficient condition to call a cluster as solid-like or liquid-like because the liquid-like cluster can also be connected frequently because of the presence of local order in liquids. Therefore an additional condition has been included that if the number of ``connections" are above some threshold, say 6 or 8 then the particles are solid-like, and if less, then they are liquid-like. Using this criterion, the solid-like particles can be distinguished from the liquid-like particles as the former will have more number of connections (coordination) than the latter. Distribution of number of connections per particle in LJ system for liquid, bcc, and fcc structures is shown in Figure \ref{fig:wolde}.

The approach of Steinhardt, ten Wolde, Frenkel, and coworkers has been further extended by Eslami {\em et al.}~\cite{eslami2017local} who have defined a local OP as follows, 
\begin{equation}
 q_l(i) =\frac{1}{N_b(i)}\sum_{j\in N_b(i)}\bf{q_l(i)}.\bf{q_l(j)}^*
\end{equation}
The $q_l(i)$ OP includes contributions from the first and second coordination shell neighbors of the particle $i$ averaged over all its neighbors.

Lenchner and Dellago proposed a variant of the Steinhardt OPs which is more accurate to determine specific crystal structures.~\cite{lechner2008accurate}. They introduced the OPs that are obtained by averaging the bond orientation orders over all neighbors,

\begin{equation}
\bar{q}_{l}(i) = \sqrt{\frac{4\pi}{2l + 1} \sum_{m=-l}^{l}|\bar{q}_{lm}(i)|^2}
\end{equation}

where $\bar{q}_{lm}(i)$ is defined as,
\begin{equation}
    \bar{q}_{lm}(i) =\frac{1}{\Tilde{N}_b(i)}\sum_{k=0}^{\Tilde{N}_b(i)}q_{lm}(k)
\end{equation}
where the index $k$ goes from 0 to all $N_b(i)$ neighbors of $i$ including itself. The advantage of this definition are - these OPs are not restricted to including only the first coordination shell but they can take into account the second shell neighbors, and the solidlike and liquidlike particles can be distinguished better due to  the decreased overlap between the OPs distributions belonging to the two states.

They have calculated the average of the probability distributions of $\bar{q}_4$ and $\bar{q}_6$ for fcc, bcc, and hcp crystals in undercooled liquid in which particles are interacting {\em via} LJ and Gaussian potential (Table \ref{fig:TAB2} and  \ref{fig:TAB3}). Due to the averaging procedure the overlap between order parameter distributions in different phases decreases. As a result sharper distinction between different phases is obtained.

\begin{center}
\captionof{table}{Global orientational order parameters for fcc, hcp, bcc, sc, and icosahedral structure.(\em{ Reprinted with the permission from AIP publisher from Ref. \citenum{lechner2008accurate})}}
\begin{tabular}{  m{4cm} m{2cm} m{2cm}   m{2cm}  m{2cm}  }
\hline \hline
& $q_4$ & $\bar{q}_4$ & $ q_6$ & $ \bar{q}_6$  \\
\hline
bcc & 0.089 988 & 0.033 406 & -0.440 526 & 0.408 018  \\ 
fcc& 0.170880 & 0.158180 & 0.507298 & 0.491385 \\ 
hcp & 0.107923 & 0.084052 & 0.445384 & 0.421813 \\ 
liq & 0.109049 & 0.031246 & 0.360012 & 0.161962\\
\hline
\end{tabular}

\label{fig:TAB2}
\end{center}

\begin{center}
\captionof{table}{Average of the distributions of q4 , $\bar{q}4 $, q6 , and $\bar{q}6$ for the bcc, fcc,and hcp crystals in  system interacting via Gaussian potential.(\em{Reprinted with the permission of AIP publisher from Ref. \citenum{lechner2008accurate})}}
\begin{tabular}{  m{4cm} m{2cm} m{2cm}   m{2cm}  m{2cm}  }
\hline \hline
& $q_4$ & ${\bar{q}}_4$ & $q_6$ & ${\bar{q}_6}$  \\
\hline
bcc & 0.085581 & 0.031728 & 0.437129 & 0.407515 \\ 
fcc& 0.155336 & 0.134388 & 0.474079 & 0.447782 \\ 
hcp & 0.109723 & 0.073369 & 0.424627 & 0.385720 \\ 
liq & 0.126950 & 0.040297 & 0.375121 & 0.158913\\
\hline
\end{tabular}
\label{fig:TAB3}
\end{center}

\begin{figure}[H]
    \centering
    \includegraphics[width=0.4 \linewidth]{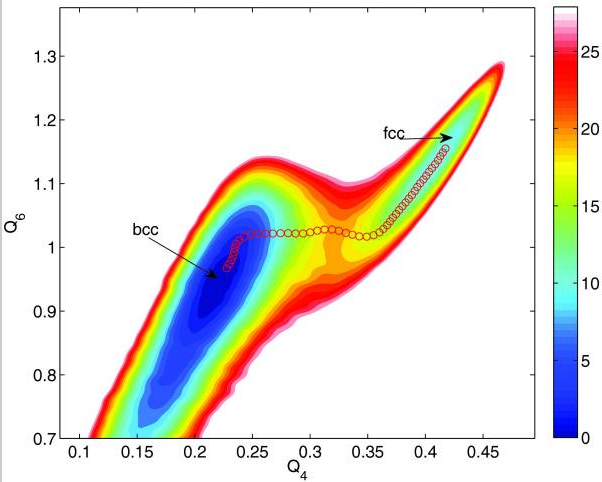}
    \caption{Free energy surface when only $Q4$ and $Q6$ were used as CVs.(\em{Reprinted with the permission of AIP publisher from Ref. \citenum{yu2014})}}
    \label{fig:tang}
\end{figure}

In 2014, Tang {\em et al.} have used as CVs the local and average Steinhardt OPs with or without supercell parameters included in the CVs definition to predict the polymorphism of Xenon crystal at high temperature and pressure.\cite{yu2014} As expected, fcc and bcc structures were obtained when $Q4$ and $Q6$ were used as CVs (see Fig \ref{fig:tang}), and along with it, the new structures including fcc with hcp stacking faults were also obtained when the supercell parameters was included in the CVs definition.

\begin{figure}[H]
    \centering
    \includegraphics[width=0.7 \linewidth]{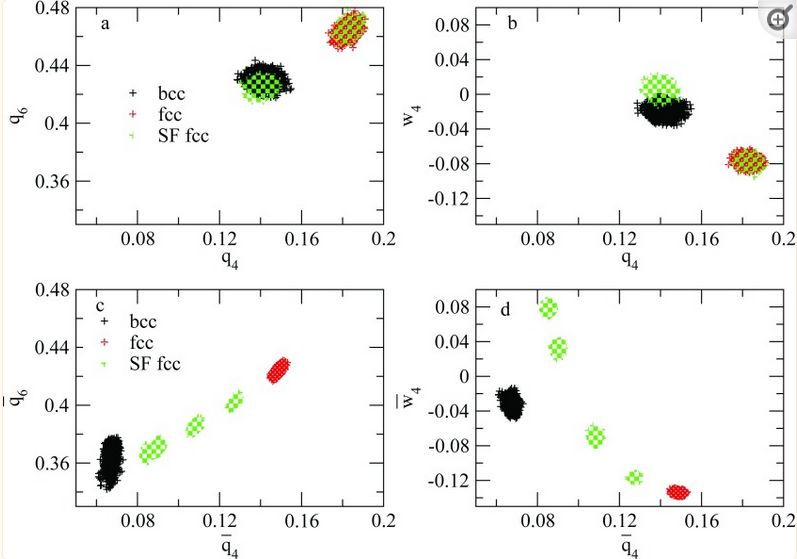}
    \caption{The local (q6, q4, and w4) and the average ($\bar{q}6$, $\bar{q}4$, and $\bar{w}4$) bond OPs of various solid forms are plotted for comparison.\em{Reprinted with the permission of AIP publisher from Ref. \citenum{yu2014}}}
    \label{fig:locavg}
\end{figure}

The structures obtained by using local and average order parameters as CV were compared (see Fig. \ref{fig:locavg}). The fcc structure with stacking faults show two types of structures with the local OPs were used as CV, and the fcc structure with stacking faults were found to split into more types of structures when the average OP was used as CV.

\begin{figure}[H]
    \centering
    \includegraphics[width=0.8 \linewidth]{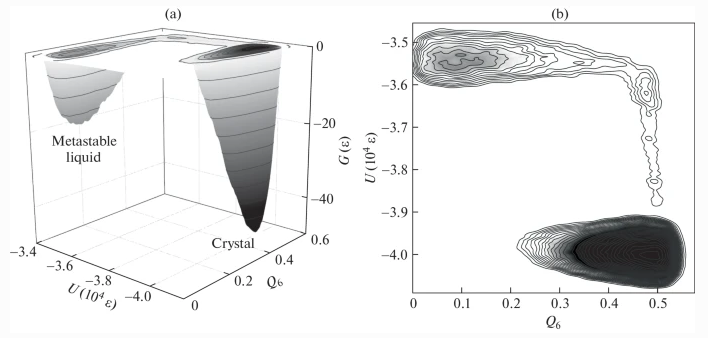}
    \caption{ (a)Free energy landscape for the system (b) Projection of the same landscape in the Q6-U space.(\em{Reprinted with the permission of Springer publisher from Ref. \citenum{rozanov2022study})}}
    \label{fig:Q6U}
\end{figure}

Very recently, Rozanov {\em et al.} has studied the phase transition of a LJ system from its metastable liquid to crystalline state using MetaD simulations.~\cite{rozanov2022study} The $Q6$ along with the systems potential energy ($U$) were used as CVs. The free energy landscape associated with this phase transformation process is shown in Figure \ref{fig:Q6U}.

The free energy barrier of crystallization at constant pressure obtained from the MetaD simulations is in good agreement with those obtained from experiments. This manifests the CVs effectiveness in sampling the actual nucleation process and reproducing the experimental observations.~\cite{rozanov2022study}

\noindent
\subsection{2.2. SOAP Kernel}

\noindent
\subsubsection{2.2.1. SOAP Kernel method}
A crucial property of a variable representing atomic environments is its invariance with respect to the basic symmetries like rotation, reflection, translation, and permutation of atoms. Steinhardt OPs are one of the invariants used to describe atomic environments. But these OPs have certain limitations.~\cite{Mickel2013} The qualitative trend of the Steinhardt OPs is influenced by the choice of neighborhood, and due to the discrete nature of neighborhood definition, the neighborhood of a particle is not a continuous function of particles coordinates. This discontinuous nature of $q_l$ leads to the lack of robustness of these OPs as structure metrics. Bartok {\em et al.} has shown that the descriptors like Steinhardt parameters are the special cases of some general approach in which the atomic environment is defined by neighbourhood density.~\cite{bartok2013representing} This approach is called Smooth Overlap Of Atomic Positions (SOAP). In the SOAP Kernel method, each atom in a given environment is defined as the sum of Gaussian functions centered on the neighborhood of an atom and including that atom itself.  These parameters fulfil the criteria of being invariant and  continuous functions of atomic coordinates. The SOAP kernel and its variants have been used in many applications for identifying crystal structures and polymorphs.~\cite{de2016comparing,Piaggi2017} \\

\noindent
\subsubsection{2.2.2 Environment Similarity CV}
\begin{figure}[H]
    \centering
    \includegraphics[width=0.85 \linewidth]{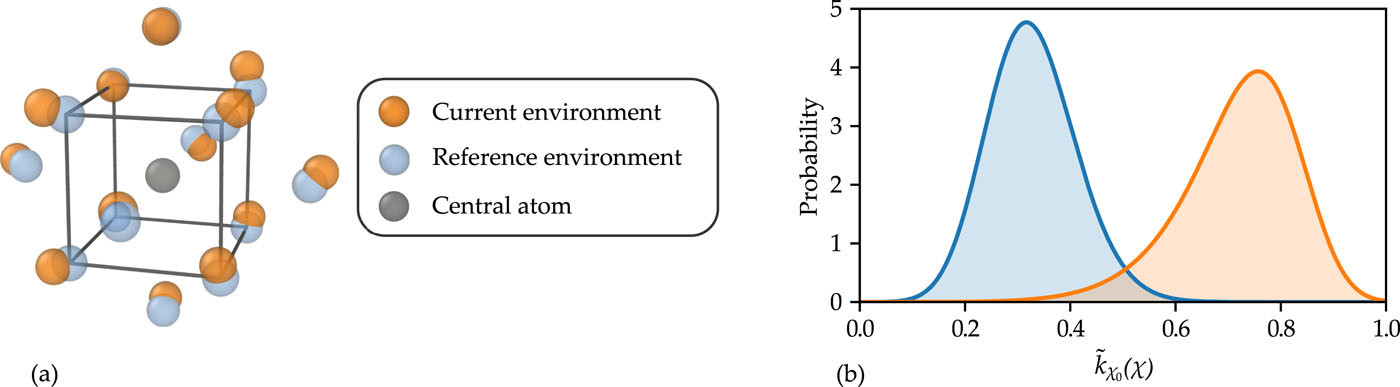}
    \caption{(a) The particles positions with respect to a central atom (black sphere) in the reference crystal environment, $\chi_0$ and in an instantaneous configuration $\chi$ are shown in grey and orange spheres, respectively.  The configuration was extracted from a trajectory of bcc sodium simulated at T = 300 K and P = 1 bar. (b) Distributions of the kernel $\bar{k}{_{\chi_0}(\chi)}$ (eq. \ref{eq:nkernel}) ) for the liquid (blue) and bcc (orange) sodium at 375 K and 1 atm pressure.(\em{Reprinted with the permission of AIP publisher from Ref. \citenum{piaggi2019calculation})}}
    \label{fig:envcv}
\end{figure}

Piaggi and Parrinello utilized a reduced definition of the SOAP kernel approach to design a CV, called environment similarity CV (Env-CV).~\cite{piaggi2019calculation} This method is based on the local ordering of $n$ neighbors ($i$) around a central atom ($\bf{r}$) in a crystalline environment (Fig. \ref{fig:envcv}).  As done in the SOAP kernel approach, the local density $\rho_{\chi}(\bf r)$ of the central atom in a given environment $\chi$ is written as the sum of Gaussian functions,
\begin{equation}
\begin{split}
\rho_\chi(\bf r) &= \sum_{i\in \chi} e^{-|{\bf r_i} - {\bf r}|^2/2\sigma^2}
\end{split}
\label{eq:rho}
\end{equation}
where ${\bf r_i}$'s are the coordinates of the neighbors relative to the central atom, and $\sigma^2$ is the variance of the Gaussian functions. The $n$ nearest neighbors positions {$\{\bf r_j^0\}$} of the central atom in a reference crystal environment ($\chi_0$) are chosen. The difference between the two environments $\chi$ and $\chi_0$ is obtained from the following integral,
\begin{equation}
\begin{split}
k_{\chi_0}(\chi) &= \int dr \rho_\chi(\bf r) \rho_{\chi_0}(\bf r)
\end{split}
\label{eq:int}
\end{equation}
where $\rho_{\chi_0}(\bf r)$ is the local density of the atom in the reference crystal environment ($\chi_0$). 

Unlike in the SOAP's actual definition~\cite{bartok2013representing}, in this approach, only the spatial part has been considered. This results in a simple analytical expression of the CV in the form of a kernel function, $k_{\chi_0}(\chi)$ which can be calculated efficiently.   
\begin{equation}
\begin{split}
k_{\chi_0}(\chi) &= \sum_{i\in \chi} \sum_{j\in \chi_0} \pi^{3/2} \sigma^3 e^{-|{\bf r_i} - {\bf r^0_j}|^2/4\sigma^2}
\end{split}
\label{eq:kernel}
\end{equation}

The kernel function in Eq. (\ref{eq:kernel}) is then normalized such that similarities between identical environments $\bar{k}_{\chi_0}(\chi_0)$ is equal to one. 
\begin{equation}
\begin{split}
\bar{k}{_{\chi_0}(\chi)} &= \frac{k_{\chi}(\chi_0)}{k_{\chi_0}(\chi_0))}\\
           &= \frac{1}{n}\sum_{i\in \chi} \sum_{j\in \chi_0} e^{-|{\bf r_i} - {\bf r^0_j}|^2/4\sigma^2}
\end{split}
\label{eq:nkernel}
\end{equation}

In a system of $N$ particles, for each 
particle ($i$ = 1,..,$N$) the kernel function,  $\bar{k}_{\chi_0}(\chi_i)$ is calculated using Eq. (\ref{eq:nkernel}). The particles having $\bar{k}_{\chi_0}(\chi_i)$ $>$ ${k_0}$, where ${k_0}$=0.5, are classified as crystalline or liquid by a continuous
and differentiable switching function ($s^O_{i}$) as follows,
\begin{equation}
\begin{split}
s_{i} &= 1-\frac{1-(\bar{k}_{\chi_0}(\chi_i)/k_0)^p}{1-(\bar{k}_{\chi_0}(\chi_i)/k_0)^q}
\end{split}
\label{eq:cv1}
\end{equation}
The variable $s_{i}$ has values in the range from 0 to 1; for atoms in the solution $s_{i}$ $\approx$ 0 while those in a perfect crystalline environment, $s_{i}$ $\approx$ 1 (Fig. \ref{fig:envcv}(b)). The parameters $p$ and $q$ control the steepness and the range of the switching function. The kernel function defined in Eq.(\ref{eq:kernel}) is in the similar spirit as that of the local metric order by Martelli,  Car, and co-workers.~\cite{martelli2018local}

The Env-CV has been successfully used in the study of phase diagram of sodium and aluminium using a multithermal-multibaric enhanced sampling simulation approach.~\cite{piaggi2019calculation} The non-rotationally invariant nature of the CV facilitates the crystallization of a defect-free crystal. Niu {\em et al.} extended the application of this CV to calculate the phase diagram of Gallium modelled using a DeepNN potential.~\cite{niu2020ab} Recently, Piaggi {\em et al} used the Env-CV to study molecular system, in particular, ice nucleation from water modelled using a deep learning based potential model. The Env-CV's use was not restricted to single-component systems crystallizations. Karmakar {\em et al.} used this CV to nucleate NaCl from its supersaturated aqueous solution using a combination of MetaD and constant chemical potential MD approaches.~\cite{karmakar2019molecular}

\subsection{2.3. Molecular ordering}
\noindent
\subsubsection{2.3.1. A generalized set of OPs}
For highly symmetric systems consisting atoms or spherical particles (colloids), Steinhardt bond OPs are found effective. However, for complex low symmetric systems extending such OPs are way difficult to execute.

To define OPs for complex crystals, preliminary information can be generated from a normal unbiased MD simulation of the crystal. An OP to define complex crystallization is to take advantage of the structural properties of the given crystal or consider the atomic coordinates relevant to a specific molecular arrangement in a crystal. A new method to design an OP to study complex crystal systems is presented here.\cite{doi:10.1021/jp052535q}

\begin{figure}[h]
    \centering
    \includegraphics[width=0.3
    \linewidth]{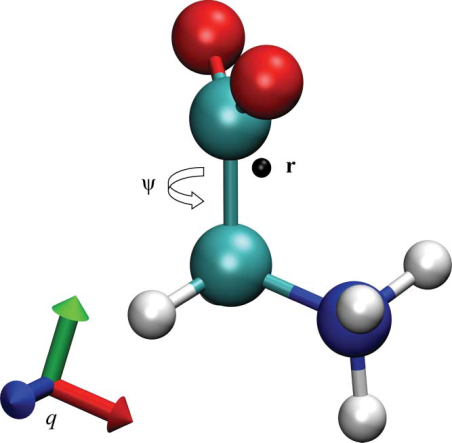}\hspace{40pt}
    \includegraphics[width=0.4
    \linewidth]{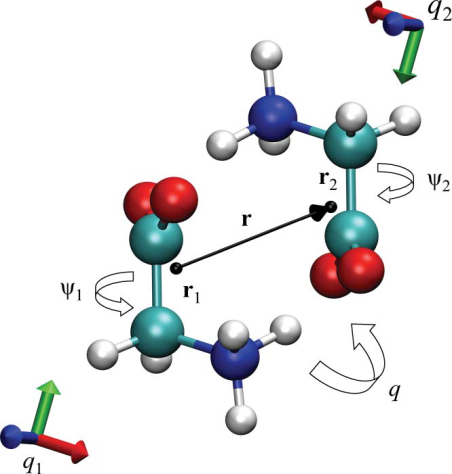}
    \caption{(a) A `point molecule' representation of glycine - $r$ is the centre of mass (COM) of the molecule, $q$ (quaternion) indicates the absolute orientation of the molecule in the reference coordinate frame (bottom left), $\psi$ is the (N-C-C-O) dihedral angle. (b) Construction of the pair distribution function - $\bf{r}$ distance vector along the COM-COM positions of the two glycine molecules projected on the first molecular coordinate frame $q1$, and $\psi_1$ and $\psi_2$ are the internal degrees of freedom that are included in the pair distribution function's definition, Eq.\ref{eq:Gffrq}. (\em{Reprinted with the permission of Springer publisher from Ref. \citenum{santiso2011general})} }
\label{fig:N1}
\end{figure}

To build OPs for molecular crystals, a generalized pair correlation function consisting of all relevant variables that represent the crystal structure is introduced by Santiso {\em et al}.~\cite{santiso2011general} Before moving ahead with pair correlation functions, it is important to understand the idea of point molecule representation (Fig. \ref{fig:N1}). In point molecule representation, the crystal system is reduced to defining -  
(i) position (the center of mass of each molecule in the crystal)
(ii) absolute orientation (containing a set of molecule-centered coordinate axes), and (iii)	the internal degrees of freedom (define the internal structure of the original molecule). The OPs are then defined as the product of the probability density functions $f$ as:

\begin{equation}
 G(r,q,{\psi},{\psi}^{'})\approx\sum_{\alpha=1}^{\infty} f_{\alpha}^r(r)f_{\alpha}^{\hat{r}}{(\hat{r})} f_{\alpha}^q(q) f_{\alpha}^{\psi}(\psi,\psi^{'})
 \label{eq:Gffrq}
\end{equation}

where $r$ is the distance between the center of mass of two molecules, ${\hat{r}}$ is the bond orientation, and the relative orientation of first molecule with respect to the second molecule is represented by $q$, and ${\psi}^{'}$ is the relative internal degrees of freedom with respect to ${\psi}$ (the internal degrees of freedom for the first molecule).  Now, we can define the pair distribution function as the probability that a molecule has an internal degrees of freedom between ${\psi}$ and ${\psi}$ + $d{\psi}$ with a neighbor with internal degrees of freedom between ${\psi}^{'}$ and $\psi^{'}$ + $d{\psi}^{'}$, relative orientation between $q$ and $q + dq$ at a position between $r$ and $r + dr$ with respect to the first molecule. The values $r$, $q$, ${\psi}$ and ${\psi}^{'}$ uniquely represent the peaks in the pair distribution function and define the crystal structure. 
In order to define an OP, there is a need to choose models for each function appearing in the probability density function, $G(r, q, \psi, \psi^{'})$ in Eq. \ref{eq:Gffrq} (Fig.\ref{fig:N1}(b)). Parameters are estimated for the models using an unbiased simulation. The distribution of distance between center of mass of the molecules, $f_{\alpha}^r (r)$ is approximated using a Gaussian function,
\begin{equation}
 f_{\alpha}^r (r) \approx \frac{1}{\sqrt{2\pi}\sigma_{\alpha}} exp[-\frac{(r-r_{\alpha})^2}{2\sigma_{\alpha}^2}]
 \label{eq:Gfr}
\end{equation}
Where $\sigma_{\alpha}$ is the standard deviation, $\alpha$ is the peak corresponding to the mean center of mass distance r$\alpha$. For each of the peaks in the pair distribution function, the distribution of bond orientation, $f_{\alpha}^{(\hat{r})} (\hat{r})$ is approximated using the Fisher distribution,
\begin{equation}
f_{\alpha}^{(\hat{r})} (\hat{r}) \approx \frac{\kappa_{\alpha}}{sinh\kappa_{\alpha}} exp(\kappa_{\alpha}\hat{r}_{\alpha}^T \hat{r})
\label{eq:Gfrh}
\end{equation}
Where $\kappa_{\alpha}$ is the concentration parameter, and $\alpha$ is the peak corresponding to the mean bond orientation $\hat{r}\alpha$. For the relative orientation distribution, $f_{\alpha}^q (q)$, 4D Bingham distribution can be used, as the orientation vectors are directionless on a 4D unit sphere. However, whatever studies have been done so far, bipolar Watson distribution can be a good approximation around each peaks in the pair distribution function $G(r, q, \psi, \psi^,)$:\\
\begin{equation}
f_{\alpha}^q (q) \approx \frac{1}{_{1}F_{1}(1/2,2,\xi_{\alpha})} exp[\xi_\alpha (q_{\alpha}.q)^2]
\label{eq:Gfq}
\end{equation}
where $\xi_{\alpha}$ is the concentration parameter, $q\alpha$ is the mean relative orientation, $1F1(1/2,2,x)$ is the confluent hypergeometric function, and the 4D dot product is denoted by the `.' symbol. Finally, the accurate model for the internal degrees of freedom is chosen on a case by case basis. When describing a crystal structure, the internal degrees of freedom that are considered are atom distances, angles, and dihedrals. For distance between two atoms,  a Gaussian distribution model is used, and for angles and dihedrals, the von Mises distribution is generally used.

\begin{equation}
\phi_{\alpha,i}^d=\sum_{j\neq i} \frac{1}{\sqrt{2\pi}\sigma_{\alpha}} exp[-\frac{(r-r_{\alpha})^2}{2\sigma_{\alpha}^2}]
\label{eq:phi1}
\end{equation}
where $i$ is the the central molecule, and $r_{ij}$ is the center-of-mass separation between molecule $i$ and its neighbors, $j$. Similar to this, one can define OPs that consider both bond distances and bond orientations,

\begin{equation}
\phi_{\alpha,i}^{bo}=\frac{1}{\sqrt{2\pi}\sigma_{\alpha}}\frac{\kappa_{\alpha}}{sinh\kappa_{\alpha}} \sum_{j\neq i} exp[-\frac{(r-r_{\alpha})^2}{2\sigma_{\alpha}^2}] exp[\kappa_{\alpha}\hat{r}_{\alpha}^T \hat{r}]
\label{eq:phi2}
\end{equation}
where $\hat r_{ij}$ represents the bond orientation vector projected onto the frame with the molecule $i$ at its centre. Additional order criteria that are sensitive to relative orientations of molecules include,
\begin{equation}
\phi_{\alpha,i}^{ro}= \frac{1}{\sqrt{2\pi}\sigma_{\alpha}}  \frac{1}{_{1}F_{1}(1/2,2,\xi_{\alpha})} \times \sum_{j\neq i} exp[-\frac{(r-r_{\alpha})^2}{2\sigma_{\alpha}^2}] exp[\xi_\alpha (q_{\alpha}.q)^2]
\label{eq:phi3}
\end{equation}
where, $F_{1}(1/2,2,\xi_{\alpha})$ is the confluent hypergeometric
function, $q_{\alpha}$ is the relative orientation, $\xi_\alpha$ is a concentration parameter, and `.' indicates a 4D dot product. Likewise, OPs that take into account a molecule's internal configuration can be defined. 

The ``local" or per-molecule and per-peak, OPs mentioned above can be used to quantify which molecule initiates the process of ordering as well as its extent. However, it is impractical for use in complex system, instead, a ``global" OP obtained by adding up either or both of the indices $i$ and $\alpha$ in Eqs. (\ref{eq:phi1})-(\ref{eq:phi3}) would be more convenient to use. 
The design and application of this order parameters is presented in the study of (i) crystallization of $\alpha$-glycine from solution which demonstrates how to cope with a nonsymmetric molecule with adaptable internal degrees of freedom, (ii) the crystallization (nucleation)~\cite{shah2011computer} of benzene from the melt, which serves as an example of how the OPs for a relatively high symmetric molecule are constructed, and (iii) solid-solid polymorph transformation of terephthalic acid.\\

\noindent
\subsubsection{2.3.2 Local crystallinity order}

Giberti, Salvalaglio, and Parrinello have developed CVs based on local crystallinity order for studying molecular crystals.~\cite{salvalaglio2012uncovering,salvalaglio2013controlling,giberti2015insight,giberti2015metadynamics,salvalaglio2015molecular} The synergistic effect of {\em local density} fluctuations and {\em molecular orientational} ordering has been embedded in the CVs definition. The advantages of this CV are two-fold: (i) It allows the system to explore the free energy surface without the prior knowledge about the global crystal symmetry, and (ii) its use can be extended to crystallization in multicomponent systems such as solution crystallization.~\cite{salvalaglio2013controlling,salvalaglio2015molecular}

The {\em global} CV is defined as the sum of individual molecules crystallinity order ($\Gamma_{i}$) in a system containing $N$ molecules,
\begin{equation}
 s=\frac{1}{N}\sum_{i=1}^{N}\Gamma_{i}   
\end{equation}

{\em Local density} of a central $i^{th}$ molecule is calculated by calculating its coordination number (CN) with respect to its neighbors, $j$. From the central atom, distances ($r_{ij}$) from its $j$ neighbors within a cut-off distance, $r_{cut}$ are calculated. To decide weather $j^{th}$ particle will be considered as neighbor of $i^{th}$ particle a switching function (Fermi), $f(r_{ij})$ is used,
 \begin{equation*}
 f(r_{ij})=\frac{1}{1+e^{a (r_{ij}-r_{cut})}}   
 \end{equation*}
The summation of $f(r_{ij})$ defines the neighbor density as:
\begin{equation}
n_{i}=\sum_{j} f(r_{ij})
\end{equation}

Another switching function, $\rho_{i}$  is defined to calculate the local density as a function of coordination number as:
\begin{equation}
\rho_{i}=\frac{1}{1+e^-b(n_i -n_{cut})}    
\end{equation}
where $a$ and $b$ are used to tune the slope of exponential functions in the switching functions. 
\begin{figure}[H]
    \centering
    \includegraphics[width=0.6 \linewidth]{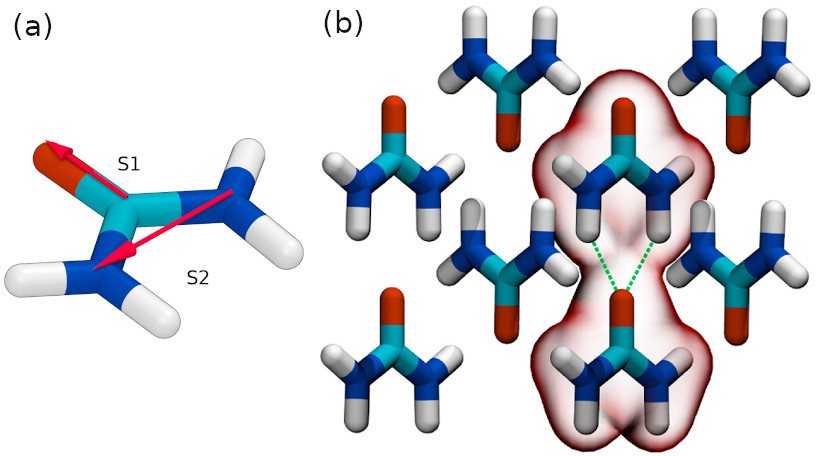}
    \caption{(a) Molecular vectors along C=O and N-N directions in a single urea molecule. (b) A representative urea crystal structure (polymorph I) extracted from WTMetaD simulation trajectory in ref.~\citenum{giberti2015insight}.(\em{Reprinted from Ref. \citenum{giberti2015insight}, Copyright (2015), with permission from Elsevier.)}}
    \label{fig:gsmac}
\end{figure}

To calculate the orientation between molecules $i$ and $j$, a function $\theta_{ij}$ which is a function of the angle between two molecular vectors is defined (Fig. \ref{fig:gsmac}(a)). Due to the effect of temperature the orientation between molecules fluctuates around an average value ($\bar{\theta}_k$). Therefore the fluctuation is expressed in terms of Gaussian functions centered around the ($\bar{\theta}_k$) between two molecular vectors,
\begin{equation}
\theta_{ij}=\sum_{k=1}^{k_{max}}e^{-\frac{(\theta_{ij}-\bar{\theta_k})^2 }{2\sigma_{k}^2}}    
\end{equation}
Here $k_{max}$ is the maximum number of angles that define the local molecular orientations. Based on these spatial and local molecular orientational OPs, the molecular crystallinity is defined as,
\begin{equation}
    \Gamma_i = \frac{\rho_i}{n_i}\sum_{j=1}^{N} f_{ij}\Theta_{ij}
\end{equation}
A CV based on the total number of crystalline
molecule is obtained by summing over the individual crystallinity orders, 
\begin{equation}
    s = \sum_{i=1}^{N} \Gamma_i
\end{equation}

In ref.~\citenum{giberti2015insight}, the fraction of molecules that are in crystalline environment has been used as a {\em global} CV, 
\begin{equation}
    s = \frac{1}{N}\sum_{i=1}^{N}\left[ \frac{\rho_i}{n_i}\sum_{j=1}^{N} f_{ij}\sum_{k=1}^{k_{max}}e^{-\frac{(\theta_{ij}-\bar{\theta_k})^2 }{2\sigma_{k}^2}}  \right]
\end{equation}

The molecular crystallinity CV has been used in my recent studies either to characterize crystal-like molecules in a multi-component system or to construct the bias potential in WTMetaD simulations studying crystallization from melt or solution.~\cite{salvalaglio2013controlling,salvalaglio2015molecular,bjelobrk2019naphthalene,bjelobrk2021solubility}\\

\noindent
\subsubsection{2.3.3. Molecular RMSD}
Inspired by the shape matching OP by Keys {\em et al.}~\cite{keys2011characterizing} and the pattern matching variables by Shetty {\em et al.}~\cite{shetty2002novel}, Duff and Peters introduced the template based polymorph specific OP which computes root mean square deviation (RMSD) between a tagged crystal molecule in a simulation system and a molecular template in a perfect crystal.\cite{duff2011polymorph} The RMSD-based OP can audibly differentiate between different polymorphs of a crystal structure.  

Though theoretically comparable to the OPs proposed by Shetty {\em et al.}, this order parameter is computationally simpler. By using a template of the crystal structure and the environment of a tagged molecule in a simulation, the method calculates the desired RMSDs. For the purpose of showcasing the new technique, Peters {\em et al.} carried out the same order parameter diagnostics as Lechner and Dellago\cite{lechner2008accurate} here for molecular crystal polymorphs. They showed that their technique is capable of differentiating between the bulk crystal structures of the three glycine polymorphs without any overlap in the order parameter distributions. Additionally, in solvated glycine crystallites, the $\alpha$, $\beta$, and $\gamma$ glycine polymorph structures may be distinguished using the local RMSD based OPs. The approach offers a broad framework that makes it simple to design OPs for a range of molecular crystals.

\begin{figure}[h!]
    \centering
    \includegraphics[width=0.5 \linewidth]{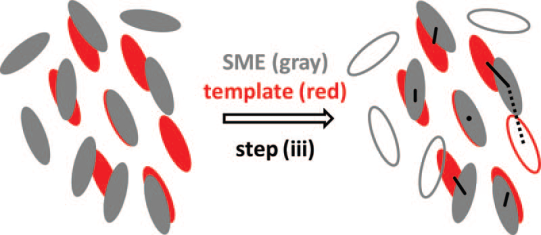}
    \caption{The grey and red ellipses represent molecules in the (SME) and in a crystal template, respectively. The core molecule in the template and the marked atom in the SME almost perfectly coincide after step (ii). The closest molecules in the SME are given the template molecules in step (iii). The closest of the two template molecules is preserved if two template molecules are assigned to the same SME molecule (as indicated by the dotted line). Unmatched SME molecules are also not kept if no template molecule is associated with a particular SME molecule. The unshaded template and SME molecules are ``deleted'' as a consequence of these criteria prior to the RMSD reduction in step (iv).(\em{Reprinted with the permission of AIP publisher from Ref. \citenum{duff2011polymorph})}}
    \label{fig:rmsd}
\end{figure}

The local molecular order is obtained by matching a tagged solute in the crystal and its simulated microenvironment (SME) comprise of neighboring solutes to its corresponding ``central'' template molecule and its neighbors present in a perfect crystal (from Cambridge Crystallographic Database or a modelled equilibrium lattice structure). If the crystal has $m$ solutes, there will be $m$ distinct templates to build the crystal structure. If one deals with $k$ number of crystal polymorphs, for each polymorph ($k$) one needs to define $m_k$ molecular templates. Fig. \ref{fig:rmsd} shows the template-matching procedure. 

Here we briefly mention the steps involved in the RMSD OPs development. To reduce the computational cost, the hydrogen atoms are not considered in the RMSD calculations, and the functional groups that can adopt degenerate configurations are replaced with non-degenerate surrogate functional groups. After this `molecular pruning' step, the RMSD OP calculation is initiated. The overall process has the following steps - (i) at first, a solute molecule is tagged and the solutes surrounding it within a cutoff radius are used to define the SME. (ii) This is followed by the rotation and translation of the central molecule in the crystal template to minimize the RMSD between the tagged and the central molecule. The same amount of rotation and translation is performed with the entire template. (iii) Subsequently, each molecule in the SME is matched with its nearest template molecule. (iv) Finally, all molecules in the SME are matched with the molecules in the crystal (polymorph) template. The final step's reduced RMSD value serves as an OP for the tagged molecule that is particular to unit-cell-member and polymorph. For each member of the polymorph $k$ unit cell, repeat steps (i-iv). Then choose the $k^{th}$ polymorph's smallest minimized RMSD value. 

Both for bulk crystallites and a small crystal in solution, the RMSD OPs can differentiate between various polymorphs with clarity. The new OPs should make it possible to simulate small molecules, something that was previously only achievable for supercooled simple liquids.

\subsection{2.3. Physical-property-based}
Recently, the use of CVs related to a system's physical properties that can be calculated experimentally has become a preferred choice for studying crystallization using simulations. This is because, in general, the value of a particular physical property is known experimentally for all the states of a system, and it can be directly used to construct a useful CV. An important aspect of using the physical property as a CV is that it does not require prior knowledge of a system's crystalline state. The physical properties that have been utilized to construct a CV are radial distribution functions, XRD peak intensities, entropy and enthalpy, and system's volume (density).\\ 

\noindent
\subsubsection{2.3.1. Radial distribution function}
\begin{figure}[!h]
    \centering
    \includegraphics[width=15cm]{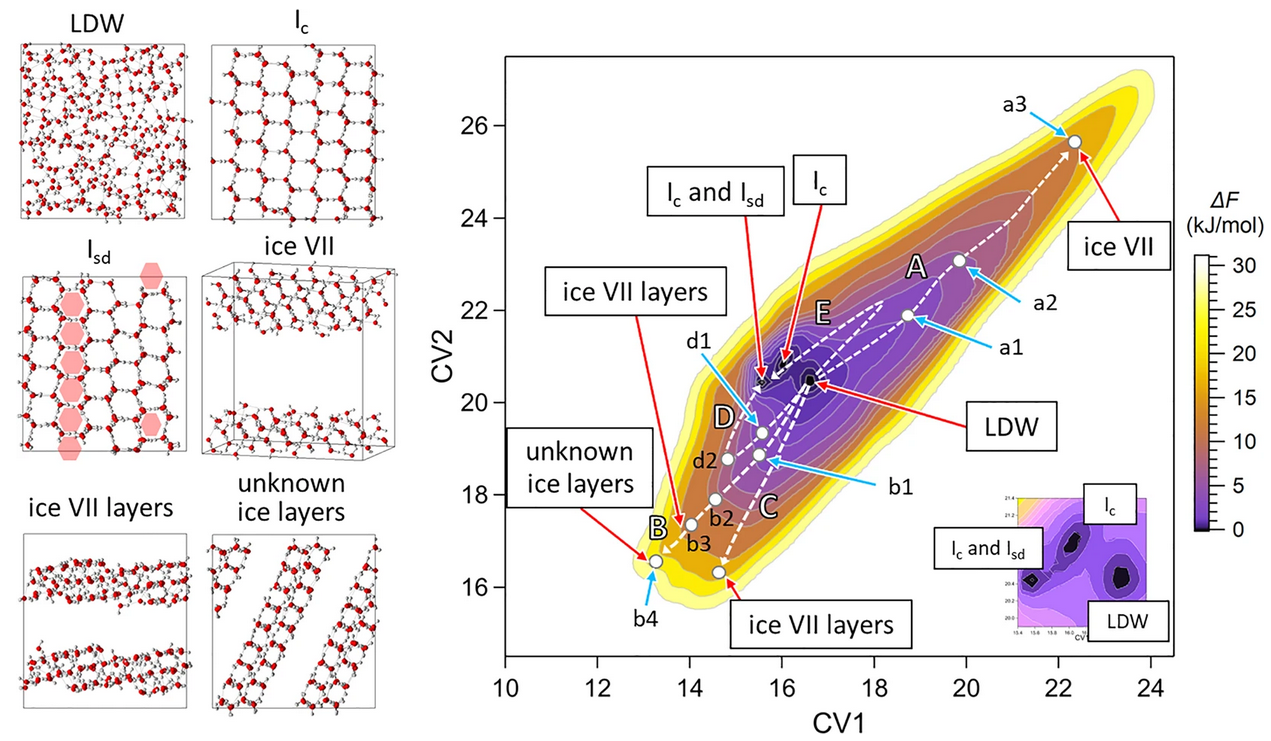}
    \caption{MetaD simulation snapshots of low density water (LDW) structures and ice structures ($I_c$ =cubic ice, $I_{sd}$= stacking disorder ice). On the right free-energy landscape is shown plotted using CV1 and CV2, where colour pallet shows free energy in kJ/mol.(\em{Reprinted with the permission of Nature from Ref. \citenum{Nada2020})}}
    \label{fig:RDF}
\end{figure}

Nada in 2020, proposed a method in which the radial distribution functions (RDFs) can be utilized as a CV for the formation of water polymorphs.~\cite{Nada2020} They performed MetaD simulations using two CVs defined as two discrete oxygen-oxygen RDFs represented by Gaussian window functions. Different polymorphs of ice such as cubic, stacking disorder~\cite{NADA1} (consists of cubic and hexagonal), high pressure ice VII, layered ice with an ice VII, and layered ice with an unknown structure were identified from the MetaD simulation trajectory (Fig. \ref{fig:RDF}).  \\

\noindent
\subsubsection{2.3.2. Entropy and Enthalpy}
The CVs discussed in sections 2.1 and 2.2 were constructed based on known crystal structures, and thus they are not effective in discovering other possible polymorphic phases of the crystal. Hence, there was a need to construct CVs that can sample the states without any prior knowledge of the crystal structure. Keeping this in mind, in 2017, Piaggi et. al. proposed the use of enthalpy and entropy surrogates as CVs.~\cite{Piaggi2017} This choice was based on two simple facts - (i) `enthalpy and entropy' - that do not predict any feature of the crystal structure a priori, and (ii) there is a trade-off between `enthalpy' and `entropy' during the crystallization which, in turn, describes the transitions between metastable states. Although `enthalpy' is easy to estimate, the `entropy' calculation is a non-trivial task. However, in the context of crystallization, we do not require an exact definition of entropy to bias the system; an approximate equation involving only two body correlations, derived from an expression where excess entropy per atom is expressed as an infinite series of terms involving multiparticle correlation functions suffices the need.~\cite{Nettleton1958}

\begin{figure}[htp]
    \centering
    \includegraphics[width=15cm]{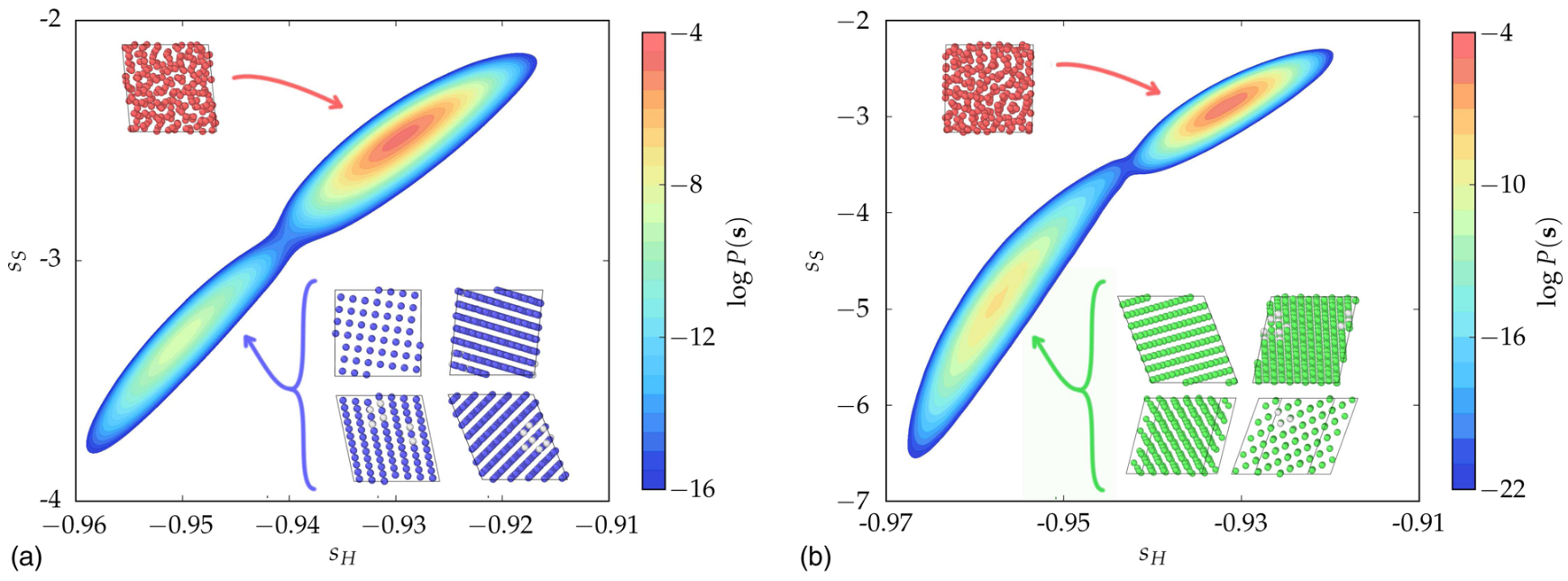}
    \caption{FES projected on the $s_H$ and $s_S$ variables for - (a) Na at 350 K and (b) Al at 800 K.(\em{Reprinted with the permission of APS Physics from Ref. \citenum{Piaggi2017})}}
    \label{fig:HS}
\end{figure}

The two CVs constructed using enthalpy and entropy are defined below:
\begin{equation}
\label{eq:HS1}
    s_H = \frac{U(R) + PV}{N}
\end{equation}
\begin{equation}
\label{eq:HS2}
     s_S = -2\pi \rho k_B \int_{0}^{\infty} [g_m(r)ln(g_m(r)) - g_m(r) + 1]r^2 dr
\end{equation}

where, $g_m(r)$ is the mollified version of the radial distribution function to ensure the function's continuity.
\begin{equation}
\label{eq:HS3}
    g_m(r) = \frac{1}{4\pi N \rho r^2} \sum_{i \neq j} \frac{1}{\sqrt{2 \pi \sigma^2}} e^{-(r-r_{ij})^2}/{2 \sigma^2}
\end{equation}
where, $\sigma$ is the broadening parameter, $r_{ij}$ is the distance between $i^{th}$ and $j^{th}$ particle, and $\rho$ is the system's density. These CVs were used to study the crystallization of Na and Al from their molten states (Fig.\ref{fig:HS}). 

Mendels {\em et al.} in 2018, extended the applicability of these CVs to study a multicomponent system, silver iodide (AgI).~\cite{Mendels2018} In this work, they have predicted the existence of an $\alpha$ phase of AgI which is stabilized by strong entropic contributions in comparison to the enthalpically-favored $\beta$ phase.

Although, these CVs were successful in predicting polymorphism in atomic crystals like Na and Al but they cannot be used for molecular crystals because molecules do not have a spherical symmetry hence they can have different orientations in space, and depending on its orientation they can exist in different polymorphic forms. The above defined CVs do not take into account  these orientations and thus are less efficient in the case of molecular crystals.

\begin{figure}[htp]
    \centering
    \includegraphics[width=8cm]{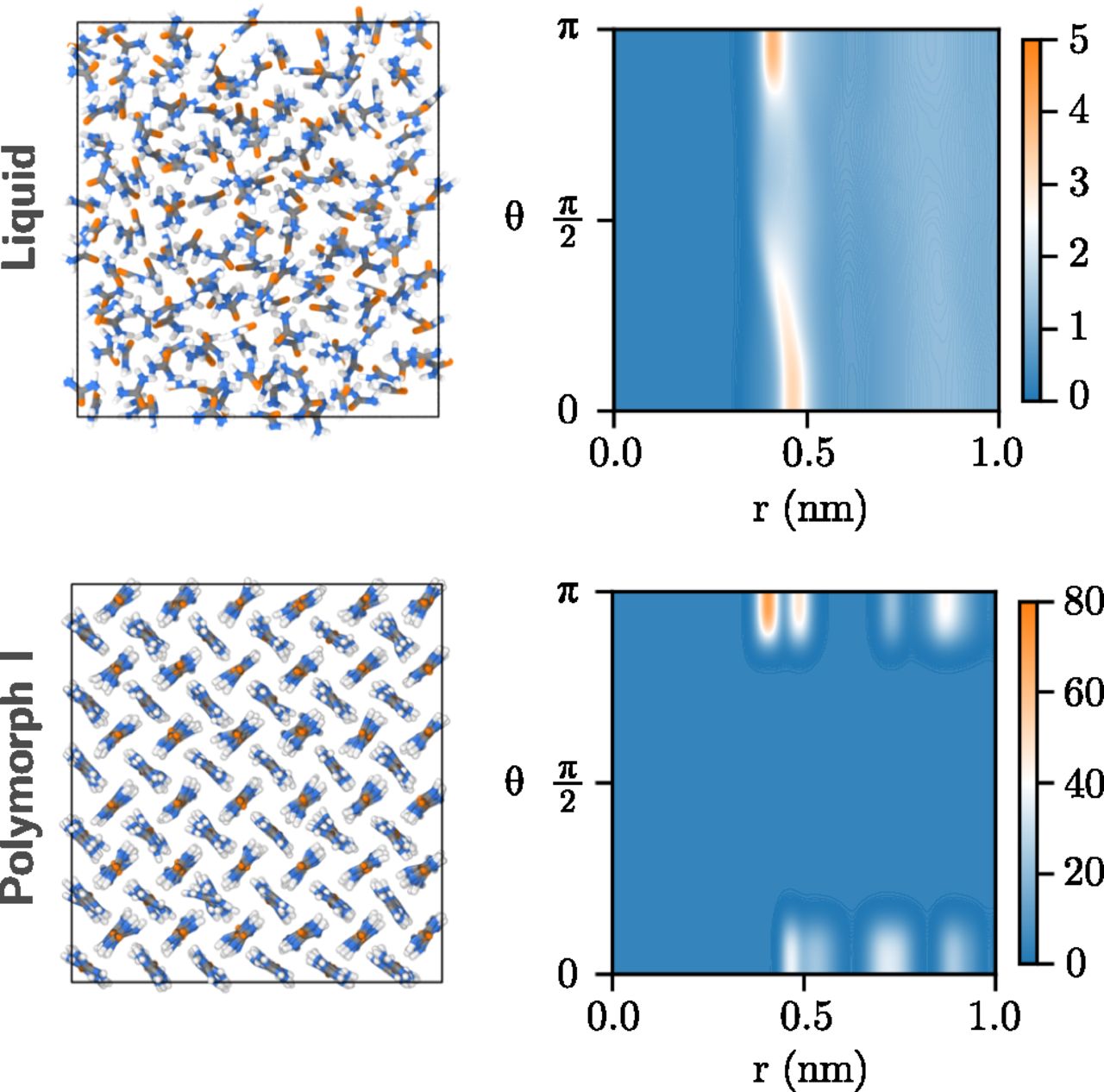}
    \caption{g(r, $\theta$) for the liquid and polymorph I of urea at 450 K with snapshots of system in both phases.(\em{Reprinted with the permission of national academy of sciences from Ref. \citenum{Piaggi2018}})}
    \label{fig:Urea}
\end{figure}

To tackle this problem, in 2018 Piaggi {\em et al.} proposed the use of orientational entropy as a CV for predicting polymorphisms in molecular crystals.~\cite{Piaggi2018} In this case, along with the spatial distances they included molecular orientations ($s_\theta$).
\begin{equation}
\label{eq:OS1}
    s_\theta = -2\pi \rho k_B \int_{0}^{\infty} \int_{0}^{\pi} [g_m(r, \theta)ln(g_m(r, \theta)) - g_m(r, \theta) + 1]r^2 sin\theta dr d\theta
\end{equation}
where, $\theta$ is the angle between two vectors $v_i$ and $v_j$ describing the orientation of molecules i and j.
\begin{equation}
\label{eq:OS2}
    \theta = arccos \left(\frac{v_i.v_j}{|v_i||v_j|}\right)
\end{equation}
In principle, at least three angles are required to define relative orientation of molecule in space, for example, Euler angles $\phi, \theta, \psi$. Hence, our function would look like $g(r, \phi, \theta, \psi)$ which is not very convenient to work with. As this function has four variables making our simulations complicated and less efficient. So, instead of taking one CV with 4 variables better alternative is to take 2 CVs $s_{\theta_1}$ and $s_{\theta_2}$ defining two different relative orientation of molecules using two angles ${\theta_1}$ and ${\theta_2}$. To understand the behaviour of $g(r, \theta)$ for liquid and solid phases we can see an example of Urea at 450 K (Fig. \ref{fig:Urea}). Here, $\theta$ represents the direction of dipole moment in urea. From the figure \ref{fig:Urea} we can observe that the liquid exhibits some structure at very short distances whereas, in polymorph I, a well-defined structure exists at long distances. One of the main characteristics of polymorph I revealed by $g(r, \theta)$ is that molecules have parallel and antiparallel dipole moments.

Further, using these CVs WTMetaD simulations were performed for urea and naphthalene at 450 K and 300 K, respectively. These temperatures are close to the melting temperatures of both substances. A large number of transitions to different crystal forms have been observed. To identify and classify the polymorphs formed during the simulation, a similarity finding strategy has been adopted in Ref. \citenum{Jones2001} and \citenum{clustering}. between two given configurations. The distance between two g(r, $\theta$)'s i.e., the divergence is calculated as
\begin{equation}
\label{eq:OS3}
    D(g_1 \parallel g_2) = \int_{0}^{\infty} \int_{0}^{\pi}\left[g_1(r, \theta)ln \frac{g_1(r, \theta)}{g_2(r, \theta)} - g_1(r, \theta) + g_2(r, \theta) \right]\times r^2 sin\theta dr d\theta
\end{equation}
This is Kullback-Leibler divergence for non-normalized functions with a minima at $g_1 = g_2$.\cite{cesa-bianchi_lugosi_2006} As $D(g_1 \parallel g_2)$ is not symmetric, it cannot be used as distance, hence distance is given by
\begin{equation}
\label{eq:OS4}
    d(g_1,g_2) = \frac{D(g_1 \parallel g_2) + D(g_2 \parallel g_1)}{2}
\end{equation}
Using hierarchical clustering and average distance between points in two clusters trajectory of urea was analyzed and different crystalline forms and liquid form were successfully distinguished.

Amodea {\em et al.} used this entropy surrogate CV along with the potential energy CV to study the effect of cooling rate during Ni$_3$Al nanoparticle freezing.~\cite{Amodeo2020} In that work, they found that by adjusting the cooling rate of Ni$_3$Al nanoparticles one can stabilize an out of equilibrium polymorph, BCC DO$-3$ structure.\\

\noindent
\subsubsection{2.3.3. Information Entropy}
In another work, Gobbo {\em et al.} used a CV based on the relative information entropy along with the Santiso and Trout's pair-distribution function based CVs to study crystallization of benzene and paracetamol.~\cite{Gobbo2018} The point molecule representation~\cite{santiso2011general} which is characterized by the position of its molecular center ($r$) and two orientation vector ($v_1$ and $v_2$) have been used. The per-molecule OP is written as, 
\begin{equation}
\label{eq:IS1}
\begin{split}
    \Gamma_{i}^{rv} = \frac{1}{n_i} \sum_{j \neq i} s(|r_i - r_j|)
    \times \sum_{\alpha = 1}^{M} \{exp[-((|r_i - r_j| - d_{\alpha})^2/2\sigma_{d_{\alpha}}^{2})]\\
    \times exp [-((\theta(v_{1i}, v_{1j}) - \theta_{1\alpha})^2/2\sigma_{\theta_{1\alpha}}^{2})]\\
    \times exp [-((\theta(v_{2i}, v_{2j}) - \theta_{2\alpha})^2/2\sigma_{\theta_{2\alpha}}^{2})]\}
\end{split}
\end{equation}
where, $M$ is the number of peaks in the joint distribution of distances and angles, $\theta_{\alpha}$ and $d_{\alpha}$ are the peak centers, and $\sigma$ is the width of the Gaussian. $s(r)$ is a switching function. 

The global average of these OPs is given as 
\begin{equation}
\label{eq:IS2}
    \Gamma^{rv_1v_2} = \frac{1}{N} \sum_i \Gamma_{i}^{rv_1v_2}
\end{equation}

These CVs discussed above have some major limitations such as - they are (i) not able to distinguish between different polymorphs, (ii) prone to have degeneracies (different configurations giving same value of CVs), and (iii) not able to describe more complex structures such as paracetamol. Hence to overcome these limitations, the authors of ref.~\citenum{Gobbo2018} used a similar approach of utilizing entropy based CVs as first proposed in ref.~\citenum{Piaggi2018}. Here instead of using the entropy surrogate as a CV, they used its distributions to differentiate the ordered state from the disordered state taking into account the long range correlations which can give better resolution of the states. To construct the CV, relevant quantities can be selected and the relative probability density $p$ is build on-the-fly and compared with the suitable reference distribution, $q$. The relative entropy is calculated using the Kullback-Leibler divergance (KLD) method\cite{Lindley1959},
\begin{equation}
\label{eq:IS3}
    KL(q \parallel p) = \int q(x) ln\left(\frac{q(x)}{p(x)}\right) dx
\end{equation}
From the above equation it is clear that the value of KLD can only be positive, and it is zero only when $q(x) = p(x)$. 
The probability density ($p(x)$) is given by 
\begin{equation}
\label{eq:IS4}
    p(x) \approx \frac{\sum_i w(x_i)g(|x-x_i|)}{\sum_j w(x_j)}
\end{equation}
where the sum runs over all elements, $g$ is the normalized Gaussian function, and $w$ is the weights. To avoid numerical instability due to very small values of $p$, the Eq. (\ref{eq:IS3}) is modified as follows,
\begin{equation}
\label{eq:IS5}
    KL\left(q \parallel \left(\frac{1}{2}p + \frac{1}{2}q\right)\right)
\end{equation}
Kernel density estimate (KDE)\cite{Silverman2018} must be evaluated on a grid to compute integrals numerically. Hence the OP takes the final form as,
\begin{equation}
\label{eq:IS6}
    {\sum_m} q(m) ln \left(\frac{q(m)}{(p(m) + q(m))/2}\right)dS
\end{equation}
$m$ runs over all the grid points, and $dS$ is the measure of the volume element associated with every grid point. 

Using this approach, one can construct CVs of increasing complexity that can help understand the crystallization of molecular systems. In ref.~\citenum{Gobbo2018}, the authors have constructed a CV,  $KL^{\hat{r}}$ where a set of distance vectors between centers of molecules is considered. For every normalized distance vector, an $azimuthal$ angle ($\theta$) and a dihedral angle ($\phi$) are calculated. The value of the CV, $KL^{\hat{r}}$ is calculated using Eq. (\ref{eq:IS6}) using uniform distribution as the reference distribution, $q$. 

\begin{figure}[ht]
    \centering
    \includegraphics[width=8cm]{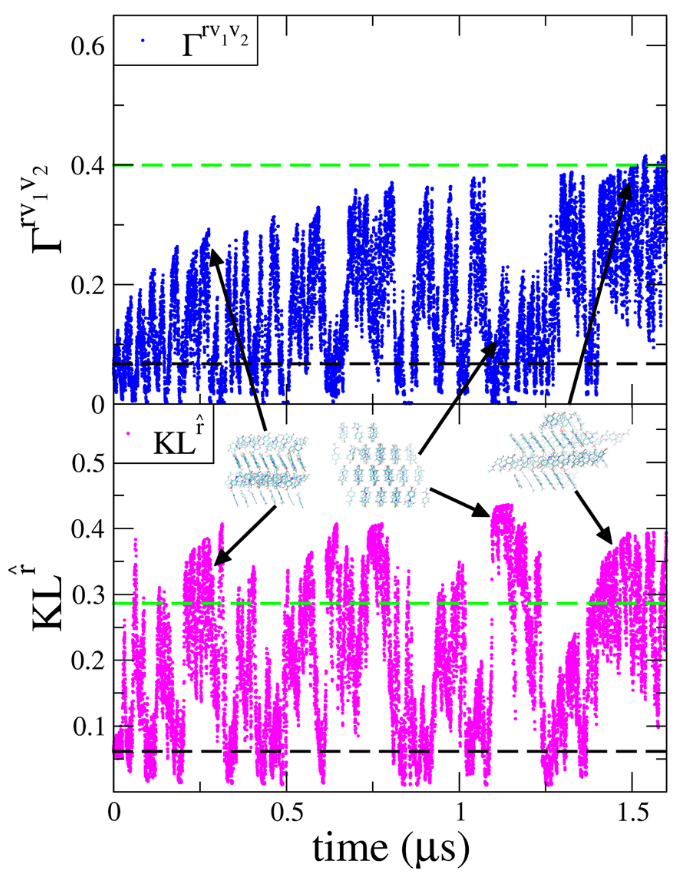}
    \caption{MetaD simulation of the paracetamol system obtained biasing both $KL^{\hat{r}}$ and $\Gamma^{rv}$.(\em{Reprinted with the permission of ACS from Ref. \citenum{Gobbo2018}})}
    \label{fig:IEn1}
\end{figure}

For benzene, a good separation between the liquid and crystal states was observed, and additionally, another ordered structure $C_2$ (possibly a polymorph) was observed as well. MetaD simulations gave two different pathways of form I crystal formation when the $KL^{\hat{r}}$ CV was used alone or used together with the pair-function based CV ($\Gamma^{rv}$). In the first pathway, orientational ordering is followed by form I crystal formation, whereas, in the second pathway, positional ordering is followed by transition to form I crystal. 

In the case of paracetamol, when both CVs, $KL^{\hat{r}}$ and $\Gamma^{rv}$ were biased, the system efficiently sampled multiple ordered and disordered states (Fig. \ref{fig:IEn1}).  A few of the ordered states resemble the form I crystal of paracetamol however, the obtained structures showed defects and unmatched lattice parameters.
\begin{figure}[htp]
    \centering
    \includegraphics[width=8cm]{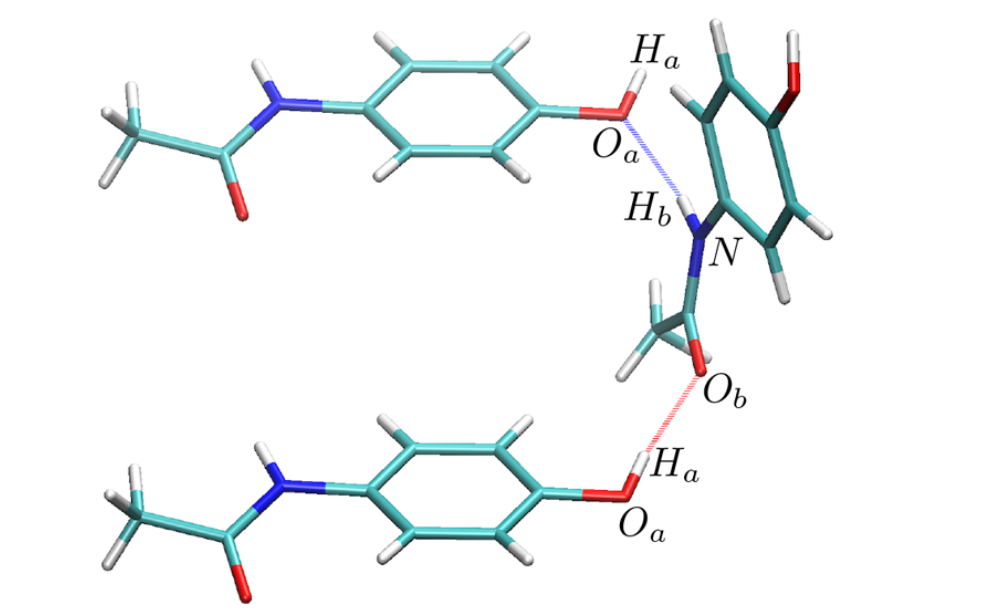}
    \caption{Example of the two unique hydrogen bonds between molecules in the form 1 crystal.(\em{Reprinted with the permission of ACS from Ref. \citenum{Gobbo2018}})}
    \label{fig:IEn2}
\end{figure}

\begin{figure}[htp]
    \centering
    \includegraphics[width=8cm]{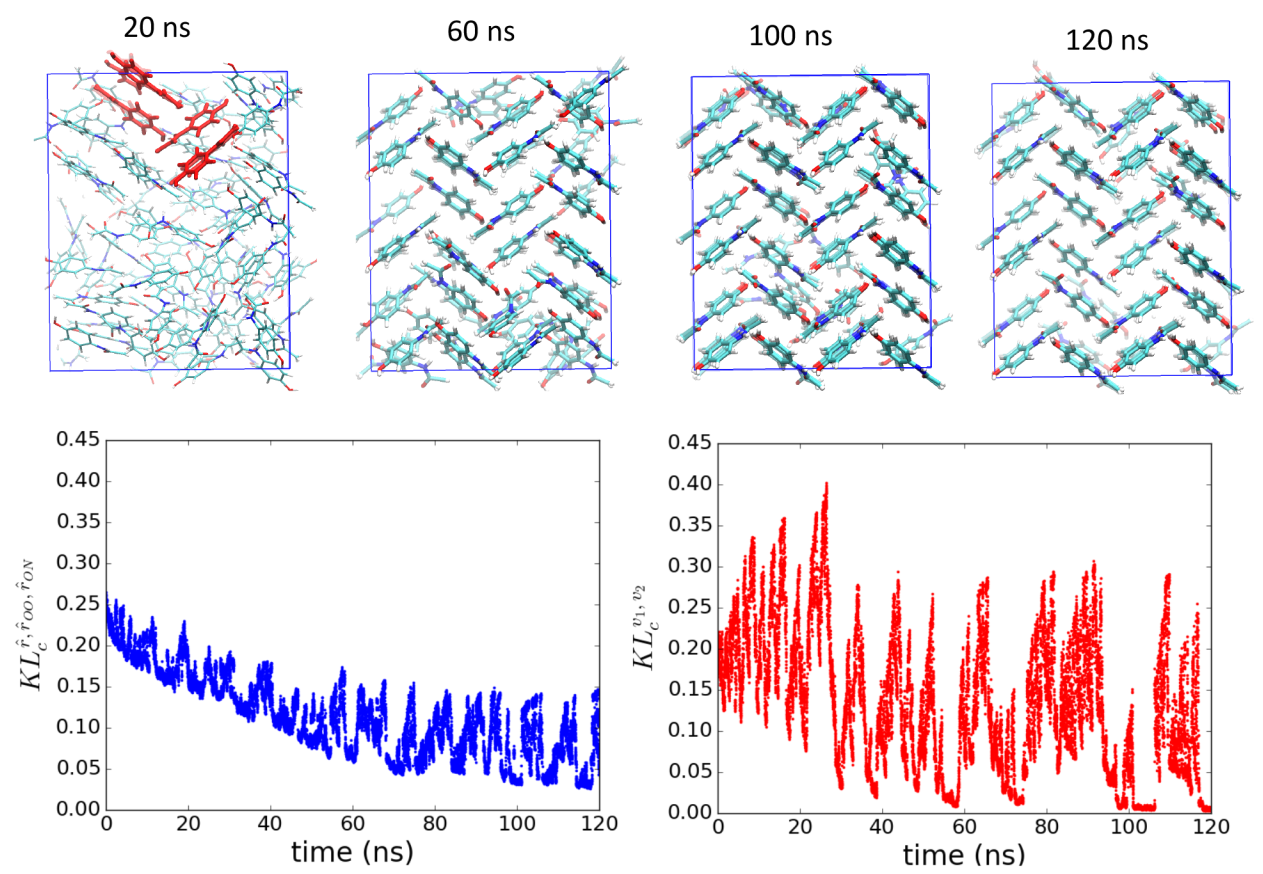}
    \caption{Nucleation trajectory obtained from a multiple walkers metadynamics simulation biasing the positional and orientational OPs, $KL_{c}^{\hat{r},\hat r_{OO},\hat r_{ON}}$ and $KL_{c}^{v_1,v_2}$.(\em{Reprinted with the permission of ACS from Ref. \citenum{Gobbo2018}})}
    \label{fig:IEn3}
\end{figure}

So to further improve the results, authors constructed OPs using only KLD framework because use of pair-function OPs was not able to define complexity of paracetamol molecules. To compensate orientational contribution which was previously described by pair-function OPs, a new KLD based OPs were constructed using orientational vectors of molecules namely $KL_{c}^{v_1}$ and $KL_{c}^{v_2}$. Instead of using them separately which will increase computational cost due to multidimentionality, they were combined to give one OP given as $KL_{c}^{v_1,v_2}$ = ($KL_{c}^{v_1}$ + $KL_{c}^{v_2}$)/2. To incorporated effect of hydrogen bonding during form I crystal formation $KL_{c}^{\hat{r}}$ was modified as $KL_{c}^{\hat{r},\hat r_{OO},\hat r_{ON}}$ = ($KL_{c}^{\hat r}$ + $KL_{c}^{\hat r_{OO}}$ + $KL_{c}^{\hat r_{ON}}$)/3. Where last two term take care of hydrogen bonds formed in from I crystal according to the fig \ref{fig:IEn2}. In metadynamics simulation it was observed nucleation event starts around 20 ns with the formation of dimers, after this system rapidly orders to form full crystal in 60ns simulation (Fig. \ref{fig:IEn3}. After analysis it was observed that $KL_{c}^{\hat r_{OO}}$ and $KL_{c}^{\hat r_{ON}}$) were the slowest evolving OPs which tells that hydrogen bond formation is the rate determining step in this nucleation process. Once the hydrogen bonds are formed in proper direction and orientation, it stabilizes the complex leading to drive the nucleation mechanism forward. Despite success of these CVs, no hydrogen bonds were found to be present in the dimeric species formed at first in the nucleation mechanism and left this mechanism unclear which leaves room for further research to understand the complexity of nucleation mechanism.\\ 

In 2020, Song {\em et al.} used the concept of Shannon information entropy based CVs to predict polymorphism in 1:1 cocrystal of resorcinol and urea using adiabatic free energy dynamics (AFED).~\cite{Song2020}\\

\noindent
\subsubsection{2.3.4. Structure factor and XRD}
Recently, Invernizzi and Niu used the concepts of structure factor and XRD-peaks, respectively, to design suitable CVs for the enhanced sampling simulation of crystallization.~\cite{invernizzi2017coarse,Niu2018} One of the most important properties of a crystal is its X-ray diffraction pattern which is easily obtained from experiments. In an XRD experiment, the scattering intensity is derived as a function of scattering vectors as follows
\begin{equation}
\label{eq:XRD1}
    I(\vec{Q}) = \sum_{i=1}^{N} \sum_{j=1}^{N} f_i(Q)f_j(Q)e^{-i\vec{Q}.(\vec{R_i} - \vec{R_j)}}
\end{equation}
where, $R_i$ and $R_j$ are the positions of $i^{th}$ and $j^{th}$ particle, $f(Q)$ is a function of magnitude of scattering vector (Q) known as the scattering form factor. 

In Ref.~\citenum{Niu2018}, the spherically averaged Debye scattering function has been used as a CV,
\begin{equation}
\label{eq:XRD4}
    I(\vec{Q}) = \sum_{i=1}^{N} \sum_{j=1}^{N} f_i(Q)f_j(Q) \frac{sin(Q.R_{ij})}{Q.R_{ij}}W(R_{ij}) 
\end{equation}
where $R_{ij}$ is the distance between the atoms $i$ and $j$. A window function $W(R_{ij})$\cite{Lorch_1970}$^-$\cite{Lin2006} is used to define a soft cutoff $R_c$ for $R_{ij}$ to avoid numerical instability while dealing with a finite-size system.

\begin{figure}[htp]
    \centering
    \includegraphics[width=10cm]{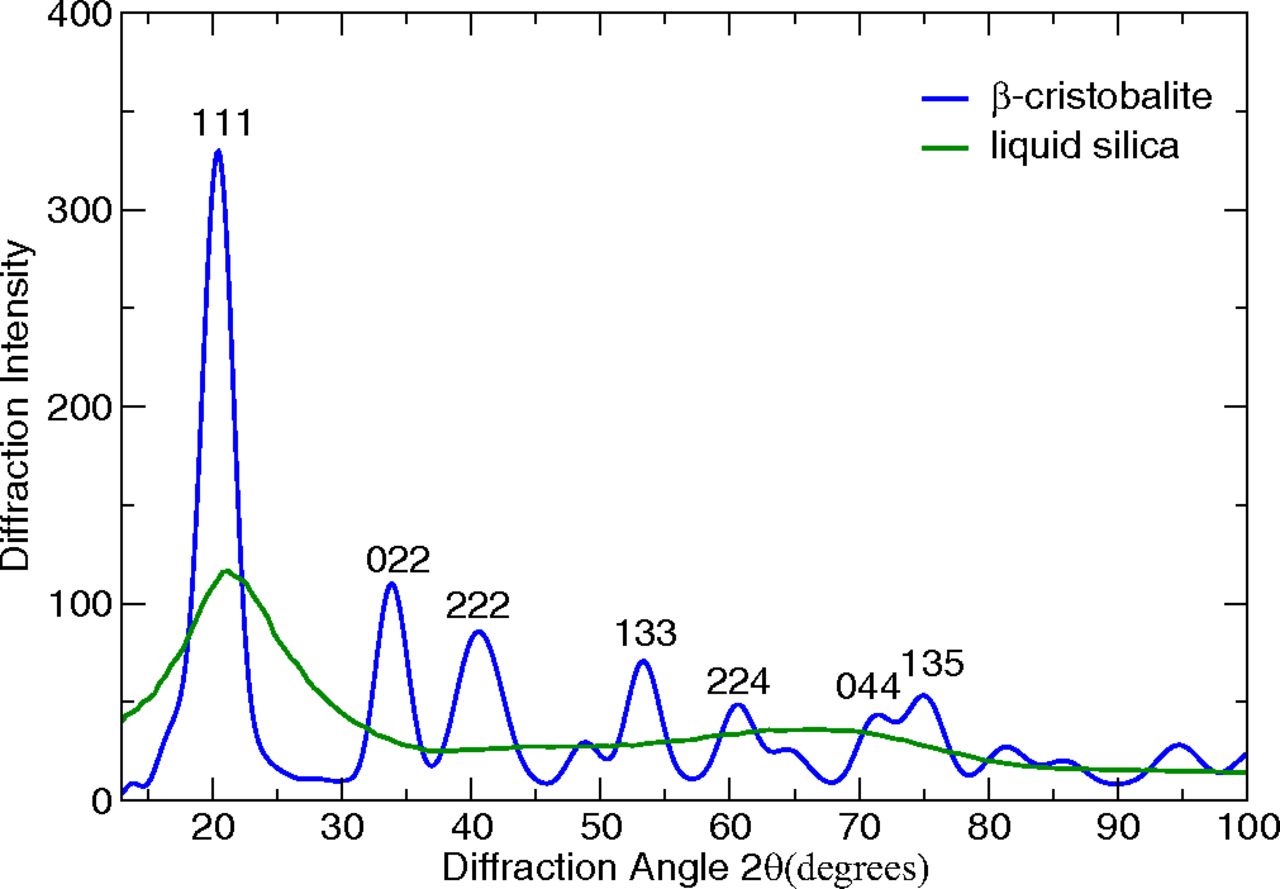}
    \caption{Simulated XRD patterns for $\beta$-cristobalite and liquid silica at 2,400 K with system containing 1536 atoms.(\em{Reprinted with the permission of ACS from Ref. \citenum{Lorch_1970}})}
    \label{fig:XRD1}
\end{figure}

In general, it is obvious to choose low-theta high intensity peaks as CVs as they provide long-range crystalline order. Ramakrishnan-Yussouff theory of crystallization also suggests to use highest peak of the structure factor as freezing order parameter.\cite{Ramakrishnan1979} Figure \ref{fig:XRD1} shows XRD patterns for $\beta$-cristobalite silica which shows most intense peak at \{111\} and \{022\} and the liquid silica where the peaks loose sharp features. The two CVs, $s_1$ and $s_2$ were therefore defined as follows,
\begin{equation}
\label{eq:XRD5}
    s_1 = I(Q_{\{111\}})
\end{equation}
\begin{equation}
\label{eq:XRD6}
    s_2 = I(Q_{\{022\}})
\end{equation}
Using these CVs, the FES for silica crystallization was obtained. Successful implementation of the XRD peak as CV has opened a new and better class of CVs which can be used for studying crystallization without any prior knowledge of the crystalline structure. Further to this development, Bonati {\em et al.} used the local structure factor as a CV to study the nucleation of silicon from its melt using a deep neural network potential for Si.\cite{Bonati2018}

\begin{figure}[htp]
    \centering
    \includegraphics[width=10cm]{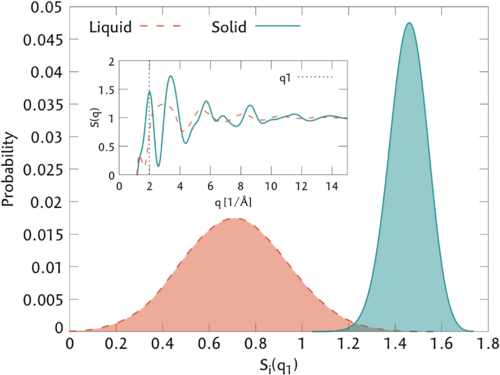}
    \caption{Distribution of the local structure factor $S_i(q_1)$ in the liquid and the solid phase.(\em{Reprinted with the  permission of APS physics from Ref. \citenum{Bonati2018}})}
    \label{fig:XRD2}
\end{figure}

The Debye formula has been modified into individual atomic contributions,
\begin{equation}
\label{eq:XRD8}
    S(q) = 1 + \frac{1}{N} \sum_{i=1}^{N} \sum_{j\neq1}^{N} \frac{sin(qr_{ij})}{qr_{ij}} = \frac{1}{N} \sum_{i}^{N} S_i(q)
\end{equation}
Here, every atom is assigned to its own structure factor $S_i(q)$ (Fig. \ref{fig:XRD2}) defined as
\begin{equation}
\label{eq:XRD9}
    S_i(q) = 1 + \sum_{j\neq1}^{N_n} \frac{sin(qr_{ij})}{qr_{ij}}
\end{equation}
Where, the sum is over all $N_n$ neighbors of atom $i$ which are in a cutoff distance of $r_c$.

In another work by Niu {\em et al.} in 2019, both XRD peak intensities and surrogate of translational entropy were used to study ice nucleation from water and their temperature dependence.\cite{Niu2019}  

In their work, the CV based on scattering peak intensities was constructed using a linear combination of seven descriptors
\begin{equation}
\label{eq:XRD10}
    s_X = s_{100} + s_{002} + s_{101} + \alpha (s_{100}^{xy} + s_{\bar{1}20}^{xy}) + \beta s_{002}^{xz} + \gamma s_{002}^{yz}
\end{equation}
here, first three peaks have high intensity, next two correspond to intensities of two main peaks of one single honeycomb bilayer which is projected into the XY plane, and the last two are the first main peak of the layers that are vertical to the honeycomb bilayer in a x-z and y-z plane respectively. $\alpha$, $\beta$ and $\gamma$ are the weights for corresponding descriptors which has the values 2, 1 and 1 respectively in this work.

The use of XRD peaks as CVs has gained popularity in recent times for the investigation of crystallization processes. There are many articles published recently where XRD peaks have been utilized as CVs.~\cite{zhang2019improving,Deng2021,Ahlawat2020,Lodesani2021,Lodesani2022} In 2021, Ahlawat {\em et al.} used XRD peaks as CVs to study phase transitions in methylammonium lead iodide (MAPbI$_3$) and formamidinium lead iodide (FAPbI$_3$).\cite{Ahlawat2020} They also found a low temperature crystallization pathway for the $\alpha$-FAPbI$_3$. In another work by Deng {\em et al.} $1^{st}$ and $2^{nd}$ XRD peak intensities were chosen as CVs to study crystallization of silica using enhanced sampling method and further combined with machine learning method to find out relationship of structure and mechanical properties of silica.\cite{Deng2021} In 2021, Lodesani {\em et al.} also utilized these CVs to study crystallization path of lithium disilicate through metadynamics simulations where they modified the equation \ref{eq:XRD4} to take only silicon atoms to calculate XRD peak intensities for the purpose of reducing computational cost.\cite{Lodesani2021} In another recent work, they  used XRD peak intensity based CVs to study  thermodynamics of silica crystallization into $\beta$-cristobalite.\cite{Lodesani2022}\\

\noindent
\subsubsection{2.3.5. Coordination number and Volume as CVs}
\begin{figure}[!h]
    \centering
    \includegraphics[width=10cm]{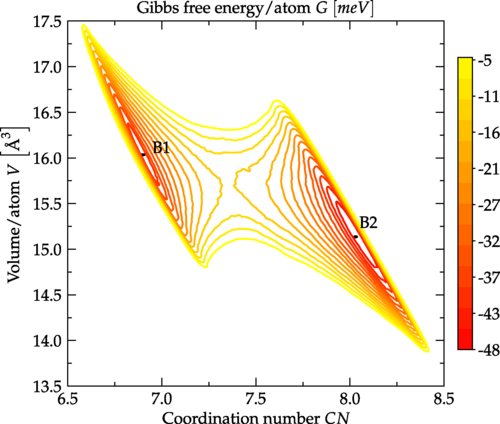}
     \caption{FES from 100 ns MetaD simulation of a 512-atom system (NaCl), using $CN$ and $V$ as CVs, at T = 300 K and P = 20 GPa.(\em{Reprinted with the  permission of APS physics from Ref. \citenum{Badin2021}})}
    \label{fig:CNV1}
\end{figure}

More recently, Badin {\em et al.} studied pressure induced B1-B2 phase transition in NaCl using metadynamics where they used coordination number ($CN$) and volume ($V$) as 2D collective variables.~\cite{Badin2021} The choice of $CN$ as CV was motivated by the generic rule of high pressure chemistry which states that pressure induced transitions are accompanied by an increase of $CN$ in the 1st coordination sphere. Also there is a significant change in the volume of the system during the pressure induced structural transition. In pressure induced phase transitions, it was a smart choice to utilize very basic properties i.e., $CN$ and $V$ as the CVs to study the B1-B2 transitions in NaCl which is accompanied by transfer of ions from the second to the first coordination shell. The average coordination number between $Na^+$ and $Cl^-$ is calculated using following switching function

\begin{equation}
\label{eq:CNV1}
    CN = \frac{2}{N} \sum_{\substack{i\in Na^+\\ j\in Cl^-}} \left (1 + \left (\frac{r_{ij}-d_0}{r_0} \right)^6 \right)^{-1}
\end{equation}

where $r_{ij}$ is the distance between $i^{th}$ cation and $j^{th}$ anion and N is the total number of of ions. $d_0$ and $r_0$ are the parameters of the switching function which can be chosen according to the need. Using these CVs free energy surface obtained for the B1-B2 transition of NaCl crystal (Fig. \ref{fig:CNV1}).

\subsection{2.4. Dimensionality reduction based}
\noindent
In the previous sections, we presented a large number of CVs that have been developed and used to carry out crystallization simulations. While they are effective on their own merits, for practical use, in enhanced sampling simulations, only a small number of such CVs can be used. However, as Russo and Tanaka\cite{russo2016crystal} pointed out, crystallization involves the ordering of multiple OPs, and to study such a process, one needs to deal with a large number of CVs/OPs. To alleviate this problem, various dimensionality reduction techniques have been used that condense a large number of CVs into one or two-dimensional ones. Here we briefly discuss some of those dimensionality methods that have been used to design CVs for crystallization simulations.\\

\noindent
\subsubsection{2.4.1. Harmonic Linear Discriminant Analysis (HLDA)}
Mendels {\em et al.} developed a method, Harmonic Linear Discriminant Analysis (HLDA) to find CVs from a set of descriptors ${\bf d(R)}$ collected from metastable states of a system. In general to construct HLDA CVs, a set of descriptors,  ${\mathbf d(R)}$ are calculated from unbiased simulations of a system's metastable phases. Subsequently, the averages and variances are used to define the `{\em between class}' (${\mathbf S_{b}} = ({\bf \mu_{A}}-{\bf \mu_{B}})({\bf \mu_{A}}-{\bf \mu_{B}})^{T} $) and `{\em within class}' ($S_{w} = (\frac{1}{\sum_{A}} + \frac{1}{\sum_{B}}) $) matrices, respectively. The highest separation between the two states is obtained by maximizing the Fischer's ratio, $\frac{W^{T}S_{b}W}{W^{T}S_{w}W}$, with respect to an $N_{d}$-dimensional projection vector, $W$. The value of $W^*$ that maximizes the Fischer's ratio is obtained as, 

\begin{equation}
W^{*}=(\frac{1}{\sum_{A}}+\frac{1}{\sum_{B}})(\mu_{A}-\mu_{B})
\end{equation}

Finally, the HLDA CV is obtained as,
\begin{equation}
s_{HLDA}(R)=W^{*T}{\bf d(R)}
\end{equation}

\begin{figure}[h]
    \centering
    \includegraphics[scale=3.64]{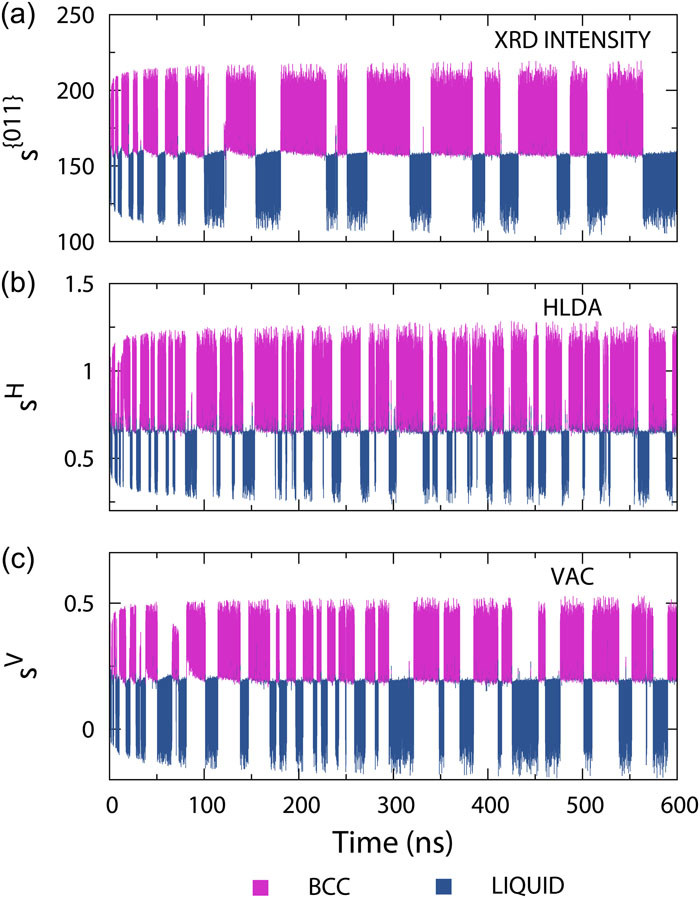}
    \caption{A comparison among three CV profiles - (a) XRD intensity CV $s^{011}$ for Na, (b) HLDA CV $s^{H}$, (c) VAC CV $s^{V}$ for Na. (\em{Reprinted with the  permission of AIP from Ref. \citenum{zhang2019improving}})}
    \label{fig:hlda1}
\end{figure}

This method has been applied in the study chemical reaction~\cite{piccini2018metadynamics} and folding of a mini-protein.~\cite{mendels2018folding} Recently, 
Zhang {\em et al.} used this approach to find out suitable CVs for crystallization of Na and Al from their molten states.~\cite{zhang2019improving} They have used a set of high intense XRD peaks (see Eq.\ref{eq:XRD4}) of crystalline Na ($d_1=\Tilde{I^{011}}$,$d_{2}=\Tilde{I^{002}}$,$d_{3}=\Tilde{I^{112}}$, and $d_{4}=\Tilde{I^{022}}$) and Al ($d_{1}=\Tilde{I^{111}}$,$d_{2}=\Tilde{I^{002}}$,$d_{3}=\Tilde{I^{022}}$, and $d_{4}=\Tilde{I^{113}}$) as descriptors to derive the HLDA CVs. Two sets of WTMetaD simulations were carried out - in the first, a single peak of the XRD was biased, and in the second set, the HLDA CV, $s^{H}$. From Fig. \ref{fig:hlda1}, it is clear that the HLDA CV, $s^H$ outperforms the single peak-based CV in sampling the solid and liquid states.

\noindent
\subsubsection{2.4.2. Time-lagged Independent Component Analysis (TICA) and Variational Approach to Conformational Dynamics (VAC)}

The time-lagged independent component analysis (TICA) linearly combines  a set of input descriptors, $\bf{d}_k(\bf{R_t})$, k = 1...$N_d$ to construct a CV as, $s_i({\bf{R}}) = \sum_{k=1}^{N_d}b_{ik} d_k$. The TICA variant developed by Pande and No{\`e} provides a way to optimally choose the expansion coefficients, $\bf{b_i}$ by solving the eigenvalue problem,
\begin{equation}
    \bf{\tilde{C}}(\tau).\bf{b_i} = \bf{\tilde{C}}(0)\lambda_i \bf{b_i}
\end{equation}
where, $\bf{\tilde{C(0)}}$ is the covariance matrix at time 0, and $\bf{\tilde{C(\tau)}}$ is the time lagged covariance matrix obtained as, $C_{mn}(\tau) = \langle r_m(0),r_n(\tau) \rangle$ where, $r_k(\tau) = d_k(\tau) - d_k$. $\lambda_i$ is the $i^{th}$ eigenvalue. The eigenvalues are arranged in descending order, and the eigenvector having the largest eigenvalue corresponding to the slowest degree of freedom is used as a CV.

Usually, TICA components are obtained from a long unbiased simulation in which the system visits metastable states multiple times as done in Refs.~\citenum{schwantes2013improvements,perez2013identification,m2017tica}. McCarty and Parrinello~\cite{mccarty2017variational} showed that a WTMetaD (biased) trajectory in which frequent transitions between the system's metastable states are obtained can also be used to obtain the TICA components. However, in the latter case, one has to obtain the scaled time from the biased simulation time,
\begin{equation}
    d\tau = e^{(-\beta V(s(\bf{R_t}) - c(t))}dt
\end{equation}

Zhang et al~\cite{zhang2019improving} used this idea to develop a VAC CV, $s^V$ which was based on the linear combination of a selected set of XRD peaks, $I^{011}, I^{002}, I^{112}$, and $I^{022}$. Both HLDA CV, $s^H$ and the VAC CV, $s^V$ exhibited improved efficiency compared to that of the single XRD peak-based CV, $s^{011}$ (Fig. \ref{fig:hlda1}). \\

\noindent
\subsubsection{2.4.3. Spectral gap optimization of order parameters (SGOOP)}
Tiwary and coworkers used the spectral gap optimization of order parameters (SGOOP) approach to construct an one-dimensional reaction coordinate (RC) to study nucleation of urea crystals.~\cite{zou2021toward} In SGOOP~\cite{tiwary2016spectral}, at first, a short MetaD simulation is performed by taking a trial CV ($f=c_{1}\psi_{1}+c_{2}\psi_{2}+...+c_{d}\psi_{d}$) to estimate the stationary density. Postprocessing optimization is performed in the space of mixing coefficients $(c_{1},c_{2},..,c_{d})$ to find out best CVs with maximum spectral gap. 
\begin{figure}[ht]
    \centering
    \includegraphics[scale=1.6]{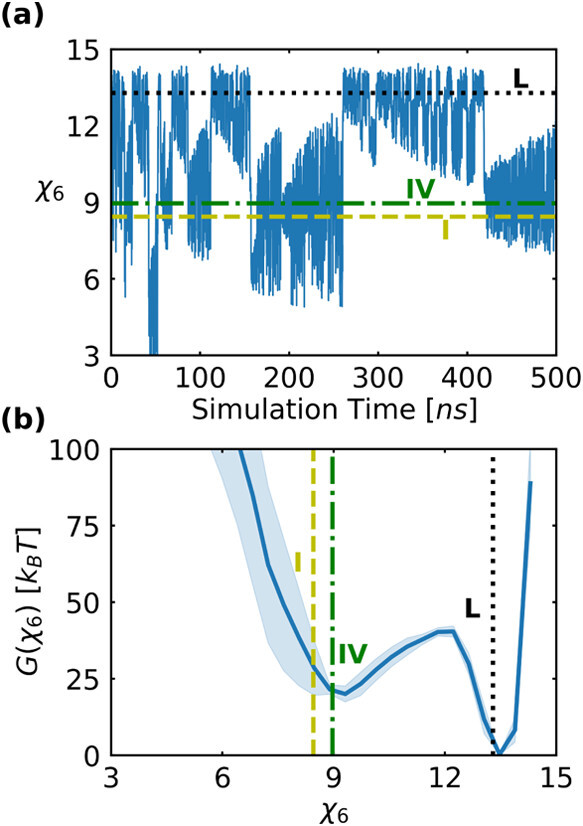}
    \caption{(a) Evolution of $\chi_{6}$ profile with time, (b) Reweighted free energy profile of WTMetaD. Polymorph I (in yellow) and IV (in green) in (a,b) figure. Multiple transition are visible from Polymorph I to IV. Error bar is shown in shaded blue color. (\em{Reprinted with the  permission of ACS from Ref. \citenum{zou2021toward}})}
    \label{fig:sgoop}
\end{figure}

In ref.~\citenum{zou2021toward}, to study urea nucleation, the RC has been defined as a linear combination of six different OPs, entropy ($S$), enthalpy($H$), coordination number, averaged angles $(\overline{\theta_1},\overline{\theta_2})$, and pair orientaional entropy $(S_{\theta_1},S_{\theta_2})$.
\begin{equation}
\chi_{6}=0.885S_{\theta_{1}}+0.330S_{\theta_{{2}}} + 0.906\overline{\theta_{1}} + 1.0\overline{\theta_{2}}-0.011N-0.017H
\label{eq:sgoop}
\end{equation}

It is clear from the coefficients of the RC defined in Eq.\ref{eq:sgoop}, the pair orientaional entropy $(S_{\theta_1},S_{\theta_2})$, specifically $S_{\theta_1}$ has maximum weight to the RC indicating its dominant role in nucleation events. The CV profile obtained from the WTMetaD simulations with the SGOOP 1d-RC shows multiple transitions of the system to various metastable states (Fig. \ref{fig:sgoop}(a)). These states correspond to different polymorphs of urea. The calculated FE profile (Fig. \ref{fig:sgoop}(b)) indicates greater stability of polymorph I than the polymorph-IV which is in agreement with the experiments.\\

\noindent
\subsubsection{2.4.4. Neural-Network-based Path Collective Variable (NN-PCV)}
Rogal {\em et al.}~\cite{rogal2019neural} have used Behler-Parrinello symmetry functions~\cite{behler2007generalized,behler2011atom} and Steinhardt parameters ($Q_l, l=6,7,8$, see section 2.1.1) as input descriptors for a feed-forward NN to obtain per-atom CVs ($q^{\alpha}_i$) corresponding to a particular crystal structure ($\alpha=$ A15, fcc, bcc, hcp, disordered structure). Subsequently, these atomic CVs were used to define the global CVs as, $Q^{bcc} = (1/N)\sum_{i=1}^{N} q_i^{\alpha}$. Finally, the path CV between two states, say $A15$ and $bcc$ is defined as,
\begin{equation}
    f({\bf{Q(r)}}) = \frac{1}{P-1}\frac{\sum_{k=1}^{P}(P-1)exp[-\lambda\lvert{\bf{Q(r)}} - {\bf{Q_k(r)}}\lvert^2]}{\sum_{k=1}^{P}exp[-\lambda\lvert{\bf{Q(r)}} - {\bf{Q_k(r)}}\lvert^2]}
\end{equation}
where, $\bf{Q(r)}$ is a point on the two-dimensional $\{Q^{A15}, Q^{bcc}\}$ space, ${\bf{Q_k}},k=1...P$ are the nodal points along the path of transformation from $A15$ to $bcc$ states, $\lvert{\bf{Q(r)}} - {\bf{Q_k(r)}}\lvert^2$ is the square distance, and $\lambda$ is a parameter. Starting from A15 phase, the value of the Path-CV increases from 0 to 1 reaching the bcc phase. Another Path-CV, $z({\bf{Q(r)}})$ perpendicular to $f({\bf{Q(r)}})$, is defined to measure the distance from the path of transformation. One can use $z({\bf{Q(r)}})$ either to construct bias or as a restraint potential.
\begin{figure}[h]
    \centering
    \includegraphics[scale=1.3]{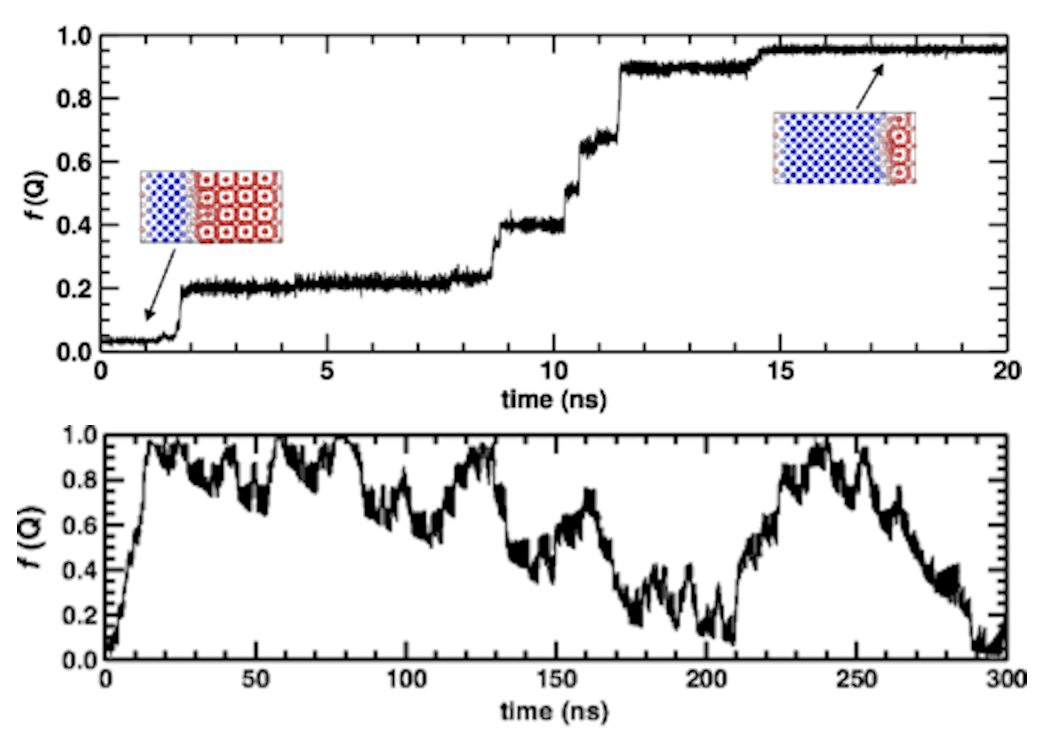}
    \caption{Path-CV, $f({\bf{Q(r)}})$ as a function of d-AFED (top) and MetaD (bottom) simulation time. The red and blue spheres indicate Mo atoms in bcc and A15 phases, respectively.(\em{Reprinted with the  permission of AIP physics from Ref. \citenum{rogal2019neural}})}
    \label{fig:hlda}
\end{figure}
This Path-CV was then used in d-AFED/TAMD and MetaD simulations to enhance the phase transition between the $A15$ to $bcc$ phases of Molybdenum and calculate the associated free energy profile (not shown).\\ 

\newpage
\noindent
\subsubsection{2.4.5. Deep-LDA}
So far we have discussed a few methods that are used to linearly combine a set of descriptors to construct low dimensional CVs. Recently, a few non-linear methods based on deep learning have been developed and used in the context of crystallization. Deep-LDA developed Bonati {\em et al.} is one such method in which a deep NN is appended with a LDA layer in the penultimate step of the dimension reduction setup.~\cite{bonati2020data} 

In the LDA method, one uses $N_d$ number of descriptors, {\bf{d(R)}} which are the functions of atom coordinates and calculate `{\em within class}',  
$\bf{S_w} = \frac{1}{2}(\sum_S + \sum_L)$, and `{\em between class}, $\bf{S_b} = (\bf{(d_S - d_L)(d_S - d_L)}^T$ matrices. Here in the context of crystallization, the subscript $S$ and $L$ refer to the `Solid' and `Liquid' states, respectively. The maximum separation between the two states is obtained by maximising the Fisher's ratio, $f(\bf{w}) = \frac{\bf{w S_b w}^T}{\bf{w S_w w}^T}$. The value of the projection vector, $\bf{w^*}$ that maximizes $f(\bf{w})$ is obtained from the generalized eigenvalue problem, $\bf{S_b w_i} = \nu_i S_w w_i$. The LDA CV finally takes the form, $s = \bf{w^*}^Td(\bf{R})$

In the Deep-LDA method, the {\bf{d(R)}} are fed to a feed-forward deep NN which results in an $N_h$ dimension output $\bf{h}$ (hidden layer). The `{\em within class}' and `{\em between class}' matrices of dimension $N_h\times N_h$ are then calculated in the $\bf{h}$ basis. The eigenvalue of the lowest eigenvector from the Fisher's generalized equation is used as the loss function to optimize the NN weights. Finally, the deep-LDA CV is obtained as, $s = \bf{w}^T\bf{h}$ 

The efficiency of a Deep-LDA CV depends on the quality of the input descriptors. As common descriptors, one can use atom coordinates, distances, and coordination numbers. However, most of these order parameters are localized and include short-range orders. To study crystallization which involves long-range ordering of molecules in a periodic lattice, Karmakar {\em et al.}~\citenum{karmakar2021collective} used the square root of the three dimensional structure factor peaks as input descriptors for Deep-LDA. 
\begin{equation}
    d({\bf{k}}) = \sqrt{s({\bf{k}})} = \frac{1}{\sqrt{N}}\bigg\rvert \sum_{i=1}^{N}e^{-i{\bf{k.R_i}}}\bigg\rvert 
    \label{eq:dk}
\end{equation}
where, $\bf{k}$ is the 3D scattering vector, and ${\bf{R_i}}, i=1...N$ are $N$ atoms coordinates. An appropriate choice of the $\bf{k}$ vectors results in the formation of a particular crystal lattice. The CV by its construction is not rotationally invariant, however, this feature favors the growth of the crystal aligned with the MD box.  

This approach has been applied in study of NaCl and CO$_2$ crystallization from their molten/fluid phase. In particular, for the case of CO$_2$, the $s{\bf{k}}$ peaks with Miller indices, $(111),(012),(121),(302),(132)$, and $(004)$ were used to describe the input descriptors (Eq. \ref{eq:dk}). The Deep-LDA CV was used in the OPES simulations to study the phase transitions. 
\begin{figure}[htb]
    \centering
    \includegraphics[scale=2.5]{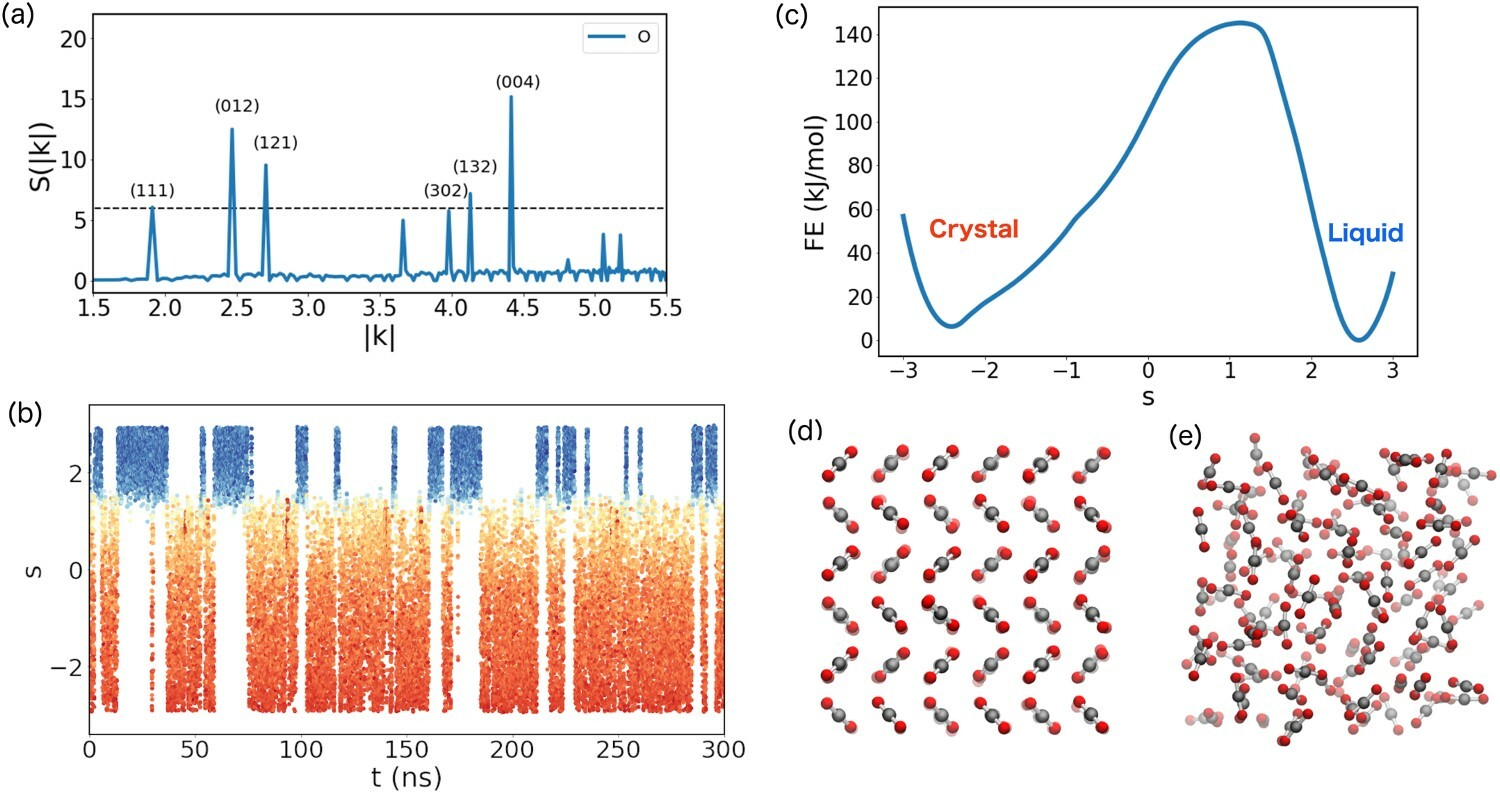}
    \caption{(a) S(k) plot of $CO_{2}$ crystal is shown with respective Miller indices, (b) The evolution of CVs profile with time obtained from OPES MetaD The color is based on value of  the miller index $(111)$ which is the first descriptors of Deep-LDA. Red color represent the crystal and blue color represent the liquid, (c) Free energy profile of $CO_{2}$ crystallisation, and (d) Crystal and liquid structure of $CO_{2}$ are shown.(\em{Reprinted with the  permission of Taylor and Francis from Ref. \citenum{karmakar2021collective}})}
    \label{fig:co2}
\end{figure}

A large number reversible transitions between the solid to liquid phase is clearly visible from Fig. \ref{fig:co2}(b) manifesting the effectiveness of the Deep-LDA CV. Similar sampling efficiency has been observed in the case of NaCl crystallization (Fig. not shown). In both cases, the Deep-LDA-based CVs gave better separation among the different states than in the single peak-based CVs. Deep-LDA based CVs give new route to efficient study of crystallization and find out the stability of crystal phase relative to its liquid phase.\\

\noindent
\subsubsection{2.4.6. DeepTICA}
In section 2.4.2, we have discussed TICA and VAC approaches to construct CVs from unbiased and biased simulation trajectories having multiple transitions between metastable states. Recently, {Bonati et al.} extended the VAC approach and developed its non-linear variant using a deep NN.~\cite{bonati2021deep}

The NN of Deep-TICA takes a set of descriptors, $\bf{d(R_t})$ and  $\bf{d(R_{t+\tau}})$ as input features and returns a set of latent variables, $\mathbf{h}_{\theta}(\bf{d(R_t)}$ and $\mathbf{h}_{\theta}(\bf{d(R_{t+\tau})}$. Subsequently, the covariance matrices are calculated using these latent variables, 
\begin{align*}
      C_{mn}(0) &= \langle h_m(t) h_n(t) \rangle\\
      C_{mn}(\tau) &= \langle h_m(t+\tau) h_n(t+\tau) \rangle
\end{align*}
The eigenvalues ($\lambda_i$) are then obtained from the generalized eigenvalue equation, 
\begin{equation}
        \bf{\tilde{C}}(\tau).\bf{b_i} = \bf{\tilde{C}}(0)\lambda_i \bf{b_i}
\end{equation}
The Deep-NN is optimized by minimizing the loss function ($\mathcal{L})$ defined as a sum of the first $N$ eigenvalues, 
\begin{equation}
    \mathcal{L} = -\sum_{i=1}^{N} \lambda_i(\theta)^2
\end{equation}
In this way, the Deep-TICA network provides as output a set of eigenfunctions that are used as Deep-TICA CVs in OPES simulations.
\begin{figure}[h]
    \centering
    \includegraphics[scale=0.18]{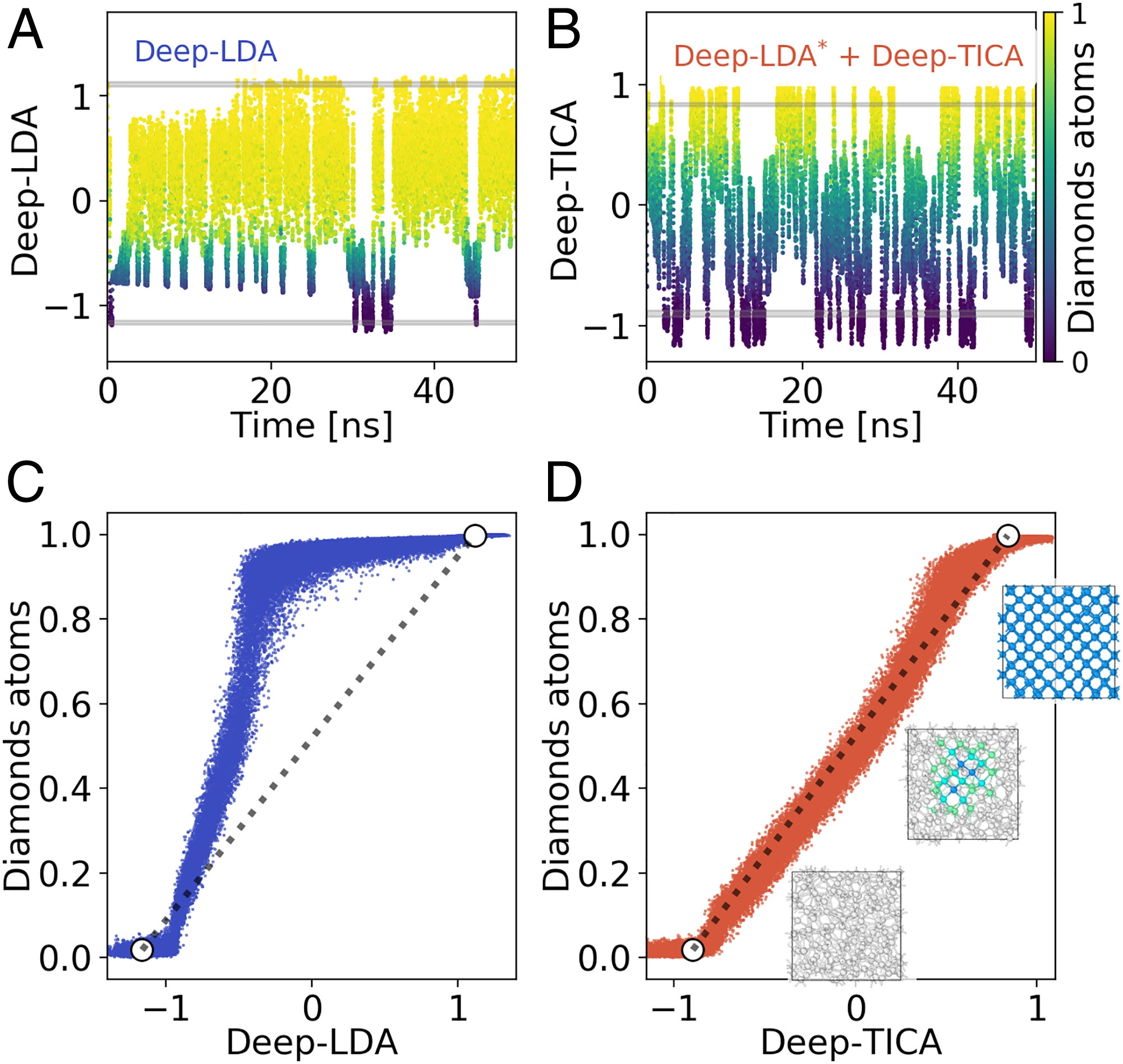}
    \caption{Comparative study between Deep-LDA and Deep-TICA driven simulation - Time evolution CV profile of (A) Deep-LDA and (B) Deep-TICA. (C) and (D) indicate the correlation between Deep-LDA and Deep-TICA CVs with the fraction of diamond atoms. White circles represent the average values of two CVs in liquid and solid phases while the dotted grey lines interpolate between them. (\em{Reprinted with the  permission of national academy of sciences from Ref. \citenum{bonati2021deep}})}
    \label{fig:deeptica}
\end{figure}

This approach has been applied to the study of prototypical processes - alanine dipeptide conformational dynamics, folding of a mini protein (chignolin), and crystallization of liquid Si. To fit the context of this review, here we discuss only the case of Si crystallization. A Deep-LDA CV was developed from a set of three-dimensional structure factor peaks, $S(\bf{k})$ (discussed in Section 2.3.4), and used in OPES simulations to sample the phase transitions. From this trajectory, the Deep-TICA CVs were obtained. Compared to the Deep-LDA CV, the Deep-TICA CV exhibited improved sampling between the two states (Fig. \ref{fig:deeptica}) manifesting the importance of incorporation of dynamical information in the development of an efficient CV.

\section{DISCUSSION}
In this review, we discussed some of the important order parameters or collective variables that have been developed and used in enhanced sampling simulations to study crystallization. The early OPs developments were based on spherical particles that were mostly used to study atomic or metallic systems. Attempts have been made to extend their application in complex multi-component materials and molecular systems. However, later studies revealed that the spherical particle-based OPs - the Steinhardt's parameters and their variants are not sufficient to capture molecular orientation in periodic crystals. The atoms' density fluctuations alone cannot fully describe the crystallinity order. The development of local molecular OPs provided a leap toward this goal, and these CVs have been used to sample the nucleation and growth of organic crystals. So far, most of the systems in which these CVs were tested consist of small, mostly rigid organic molecules. Their application in large flexible organic crystals is, however, scarce. This is due to the fact that for a large flexible molecular system, one needs to define a large number of CVs that describe molecular ordering. In ES simulation, one cannot use so many CVs, and in fact, the use of more than three variables is already a tedious task. Dimensionality reduction-based methods are useful in such a context. In this review, we have briefly discussed a few linear and non-linear (NN) methods that can help construct low-dimensional CVs from a large set of descriptors. It is important to note that the success of any dimensionality reduction-based CV development method relies on the quality of the input descriptors, and we have seen in a few examples that the linear combinations of either XRD peak intensities~\cite{zhang2019improving,niu2019temperature,Ahlawat2020} or entropy-based descriptors~\cite{zou2021toward} are found effective, while in another set of examples, the three-dimensional structure factor-based CVs were non-linearly combined to develop efficient NN CVs.~\cite{karmakar2021collective,bonati2021deep} 

Despite the enormous success of the above-mentioned approaches, the development of effective CVs for large fluxional molecules such as active pharmaceutical ingredients and biomolecules (peptides) is far from reality. The large conformational space intrinsic to these systems makes it challenging to design efficient CVs. ML-based dimensionality reduction methods~\cite{rogal2019neural,bonati2020data,karmakar2021collective,bonati2021deep} with RMSDs as effective descriptors can be tested for such systems. Among other possibilities, one can combine local atoms contacts combined with systems properties such as configurational entropy and enthalpy as possible CVs. The multicanonical approaches~\cite{piaggi2019calculation,invernizzi2020unified,bonati2021deep} along with NN-based CVs~\cite{sultan2018automated,rogal2019neural,behler2007generalized,chen2018collective,fulford2019deepice,bonati2020data,bonati2021deep} may open up new possibilities for the study of phase transitions of complex systems.\\

\noindent
{\bf 6. AUTHOR INFORMATION}\\
\noindent
{\bf Corresponding Author*}\\ 
{\bf Tarak Karmakar} - 
Department of Chemistry,
Indian Institute of Technology, Delhi,
Hauz Khas, New Delhi - 110016;
ORCID: 0000-0002-8721-6247;
E-mail: tkarmakar@chemistry.iitd.ac.in\\

\noindent
{\bf Authors}\\
{\bf Neha} - Department of Chemistry,
Indian Institute of Technology, Delhi,
Hauz Khas, New Delhi - 110016\\
\noindent
{\bf Vikas Tiwari} - Department of Chemistry,
Indian Institute of Technology, Delhi,
Hauz Khas, New Delhi - 110016\\
\noindent
{\bf Soumya Mondal} - Department of Chemistry,
Indian Institute of Technology, Delhi,
Hauz Khas, New Delhi - 110016\\
\noindent
{\bf Nisha Kumari} - Department of Chemistry,
Indian Institute of Technology, Delhi,
Hauz Khas, New Delhi - 110016\\

\noindent
{\bf Notes:}\\ 
The authors declare no competing financial interest.\\
{$^\dag$ Neha, VT, SM, and NK have contributed equally to the review article.}\\

\noindent
{\bf 7. ACKNOWLEDGEMENTS}\\
Neha, SM, NK thank IIT Delhi for institute Ph. D. fellowship, and VT acknowledges Prime Minister Research Fellowship. TK thanks IIT Delhi seed grant for financial support. 

\bibliography{ref}{}

\end{document}